\documentclass[aps,prb,preprint,superscriptaddress]{revtex4}
\usepackage{graphicx}

\begin{document}
\title{Electronic structure of Bi\textit{Me}O$_3$ multiferroics and related oxides}

   \author{J. A.~McLeod}
   \affiliation{Department of Physics and Engineering Physics, University of Saskatchewan, 
    116 Science Place, Saskatoon, Saskatchewan S7N 5E2, Canada}
	\email{john.mcleod@usask.ca}

   \author{Z. V.~Pchelkina}
   \affiliation{Institute of Metal Physics, Russian Academy of
    Sciences-Ural Division, 620041 Yekaterinburg GSP-170, Russia}
   \email{pzv@ifmlrs.uran.ru}

   \author{L. D.~Finkelstein}
   \affiliation{Institute of Metal Physics, Russian Academy of
    Sciences-Ural Division, 620041 Yekaterinburg GSP-170, Russia}

   \author{E.Z.~Kurmaev}
   \affiliation{Institute of Metal Physics, Russian Academy of
    Sciences-Ural Division, 620041 Yekaterinburg GSP-170, Russia}

   \author{R. G.~Wilks}
   \affiliation{Department of Physics and Engineering Physics, University of Saskatchewan, 
    116 Science Place, Saskatoon, Saskatchewan S7N 5E2, Canada}

   \author{A.~Moewes}
   \affiliation{Department of Physics and Engineering Physics, University of Saskatchewan, 
    116 Science Place, Saskatoon, Saskatchewan S7N 5E2, Canada}

   \author{I. V.~Solovyev} 
    \affiliation{Institute of Metal Physics, Russian Academy of
    Sciences-Ural Division, 620041 Yekaterinburg GSP-170, Russia}
    \affiliation{Computational Materials Science Center, National Institute for Materials Science, 1-2-1 Sengen, Tsukuba, Ibaraki 305-0047, Japan}

   \author{A. A.~Belik}
   \author{E.~Takayama-Muromachi}
   \affiliation{International Center for Materials Nanoarchitectonics (MANA), National Institute for Materials Science (NIMS), 1-1 Namiki, Tsukuba, Ibaraki 305-0044, Japan}

\date{\today}

\begin{abstract}
We have performed a systematic study of the electronic structures of Bi\textit{Me}O$_3$ (\textit{Me} = Sc, Cr, Mn, Fe, Co, Ni) series by soft X-ray emission (XES) and absorption (XAS) spectroscopy. The band gap values were estimated for all compounds in the series. For BiFeO$_3$ a band gap of $\sim$0.9 eV was obtained from the alignment of the O K$_{\alpha}$ XES and O \textit{1s} XAS. The O \textit{1s} XAS spectrum of BiNiO$_3$ indicates that the formation of holes is due to a Ni$^{2+}$ valency rather than a Ni$^{3+}$ valency. We have found that the O K$_\alpha$ XES  and O \textit{1s} XAS of Bi\textit{Me}O$_3$ probing partially occupied and vacant O \textit{2p} states, respectively, are in agreement with the O \textit{2p} densities of states obtained from spin-polarized band structure calculations. The O K$_{\alpha}$ XES spectra show the same degree of Bi \textit{6s}---O \textit{2p} hybridization for all compounds in the series. We argue herein that the stereochemical activity of Bi \textit{6s} lone pairs must be supplemented with inversion symmetry breaking to allow electric polarization. For BiMnO$_3$ and BiFeO$_3$, two cases of multiferroic materials in this series, the former breaks the inversion symmetry due to the antiferromagnetic order induced by particular orbital ordering in the highly distorted perovskite structure and the latter has rhombohedral crystal structure without inversion symmetry.
\end{abstract}

\maketitle

\section{\label{itro} Introduction}
Multiferroics, discovered almost 50 years ago \cite{smolenskii_61, smolenskii_63}, are materials that simultaneously possess two or three degrees of freedom: (anti)ferromagnetism, (anti)ferroelectricity, and/or ferroelasticity, which allow both charge and spin to be manipulated by applied electrical and magnetic fields \cite{scott_07, spaldin_05}. These materials are promising for various technological applications such as information storage, spintronics, and sensors. There are many different classes of multiferroic systems known today, for instance: the \textit{R}MnO$_3$ family (\textit{R} = Dy, Tb, Ho, Y, Lu, etc.), the \textit{R}Mn$_2$O$_5$ family (\textit{R} = Nd, Sm, Dy, Tb), and the Bi\textit{Me}O$_3$ family (\textit{Me} = Mn, Fe). These materials have complex structures with many atoms per formula unit, and more than one formula unit per unit cell. The large number of interatomic interactions makes distinguishing the mechanisms responsible for multiferroicity a challenging task. The origin of these phenomena and the nature of the coupling between the magnetic, electric, and structural order parameters are not well understood. Despite many electronic structure calculations (see review article \onlinecite{ederer_05} and references therein) only few experimental spectra for selected compounds (YMnO$_3$ and BiFeO$_3$) have so far been obtained \cite{higuchi_08, kang_05}. In order to get deeper insight into the nature of multiferroicity we have performed a systematic study of electronic structure of the perovskite-like multiferroics (BiFeO$_3$ and BiMnO$_3$) and the Bi\textit{Me}O$_3$ (\textit{Me} = Sc, Cr, Co, Ni) related compounds using synchrotron excited soft X-ray emission and absorption spectra. The experimental results are compared with specially performed electronic structure calculations.

The paper is organized as follows: the details of sample preparation and X-ray measurements are presented in Section II. The crystal structure of different compounds and their basic properties are summarized in Section III. Results of X-ray measurements for the whole series of Bi\textit{Me}O$_3$ (\textit{Me} = Sc, Cr, Fe, Co, Ni) compounds and comparison with electronic structure calculations are collected in Section IV. 
  
\section{\label{exp} Experimental details}

All the samples were synthesized using a high-pressure high-temperature method. Starting mixtures were placed in Au capsules and treated at 6 GPa in a belt-type high-pressure apparatus at different temperatures (heating rate was about 140 K/min). After heat treatment, the samples were quenched to room temperature (RT), and the pressure was slowly released. BiCrO$_3$ was prepared from mixtures of Bi$_2$O$_3$ (99.99 \%) and Cr$_2$O$_3$ (99.9 \%) with an amount-of-substance ratio of 1:1 at 1653 K for 60-70 min \cite{belik_08}. BiMnO$_3$ was prepared from mixtures of Bi$_2$O$_3$ and Mn$_2$O$_3$ with an amount-of-substance ratio of 1:1 at 1383 K for 60-70 min \cite{belik_mn_07}. Single-phased Mn$_2$O$_3$ was prepared by heating commercial MnO$_2$ (99.99 \%) in air at 923 K for 24 h. For the preparation of BiScO$_3$, stoichiometric mixtures of Bi$_2$O$_3$ and Sc$_2$O$_3$ (99.9 \%) were dried at 873 K for 8 h and then treated at 1413 K (at 6 GPa) for 40 min \cite{belik_06}. 

For the preparation of BiFeO$_3$, stoichiometric mixtures of Bi$_2$O$_3$ and Fe$_2$O$_3$ (99.9 \%) were first annealed at ambient pressure at 1073 K for 2 h. This procedure gave a mixture of BiFeO$_3$ (about 70 weight \%), Bi$_{25}$FeO$_{39}$, and Bi$_2$Fe$_4$O$_9$. The resulting mixture was treated at 1273 K (at 6 GPa) for 1 h. After the high-pressure treatment, single-phased BiFeO$_3$ was obtained.
BiCoO$_3$ was synthesized from stoichiometric mixtures of Bi$_2$O$_3$, Co$_3$O$_4$ (99.9 \%), and KClO$_3$ at 1483 K (at 6 GPa) for 1 h, and BiNiO$_3$ from Bi$_2$O$_3$, NiO (99.9 \%), and KClO$_3$ at 1483 K (at 6 GPa) for 1 h. The resulting samples were reground, washed with water, dried, and re-pressed at about 1 GPa at room temperature.

The X-ray emission spectra (XES) were measured at beamline 8.0.1 at the Advanced Light Source (ALS) at the Lawrence Berkeley National Laboratory \cite{jia_95}. The X-ray absorption spectra (XAS) were measured at the Spherical Grating Monochromator beamline at the Canadian Light Source (CLS) at the University of Saskatchewan \cite{regier_07}. The O \textit{1s} XAS spectra were measured in the total fluorescence yield  (TFY) mode, which provides more bulk sensitivity than electron yield methods do. The O K$_{\alpha}$ XES was excited near the O \textit{1s} ionization threshold to suppress the high-energy satellite structure. The instrumental resolving power (E/$\Delta$E) was approximately 10$^3$ for the XES measurements and approximately 5$\times 10^3$ for the XAS measurements. 

\section{\label{sec:structure} Crystal structure and physical properties of Bi\textit{Me}O$_3$ compounds}

Of all the multiferroic compounds, the most extensive study has been devoted to the perovskite-type and related oxides. 
Below we give a short summary of crystal structure and basic physical properties of the Bi\textit{Me}O$_3$ (\textit{Me} = Sc, Cr, Mn, Fe, Co, Ni) compounds. The space groups and lattice constants for the whole series are collected in Table~\ref{tab_struct}. The selected bond lengths are shown in Tables~\ref{sc_cr_mn_dist}, \ref{fe_dist} and \ref{ni_dist}. 

The structure of BiScO$_3$ has monoclinic symmetry with the space group \textit{C2/c}~\cite{belik_06}. BiScO$_3$ is nonmagnetic and most of the interest in this compound is in the ferroelectric properties of the BiScO$_3$-PbTiO$_3$ system. For the well known piezoelectric material Pb(Zr,Ti)O$_3$ (PZT) the Curie temperature T$_C$ at the morthotropic phase boundary between the rhombohedral and tetragonal ferroelectric state is 386$^\circ$ C. It has been shown that for BiScO$_3$-PbTiO$_3$ the T$_C$ reaches 450$^\circ$ C~\cite{bspt} while the piezoelectric coefficients in bulk are comparable to those of commercial PZT. 

BiCrO$_3$ crystallizes in an orthorhombic structure above 420 K with the space group \textit{Pnma} and lattice parameters  $a$ = 5.54568(12) \AA, $b$ = 7.7577(2) \AA, $c$ = 5.42862(12) \AA~at 490 K~\cite{belik_08}. A structural phase transition from an orthorhombic to a monoclinic structure occurs at 420 K \cite{belik_08, sugawara, niitaka, belik_07}. Between 420 K and 7 K  BiCrO$_3$ has a monoclinic structure with the space group \textit{C2/c} and $a$ = 9.4641(4) \AA, $b$ = 5.4790(2) \AA, $c$ = 9.5850(4) \AA, $\beta$ = 108.568(3)$^\circ$ at 7 K~\cite{belik_08}. A long-range G-type antiferromagnetic order with weak ferromagnetism develops below T$_N$ = 109 K and does not change down to 7 K~\cite{belik_08}. Four anomalies of magnetic origin were observed at 40, 75, 109 and 111 K~\cite{belik_07}. Magnetic moments of Cr$^{3+}$ ions were found to align along the monoclinic $b$ axis in a similar manner to the direction of magnetic moments of Mn$^{3+}$ in BiMnO$_3$~\cite{montanari_07, moreira_02}. The magnetic structure of BiCrO$_3$ was first predicted from \textit{ab initio} electronic structure calculations \cite{hill_02}.

\begin{table*}[!htb]
\caption{Crystal structure and lattice parameters for Bi\textit{Me}O$_3$ compounds.}
\label{tab_struct}
\begin{tabular}{|l|c|c|c|c|c|c|}
\hline
            &BiScO$_3$      & BiCrO$_3$ & BiMnO$_3$ & BiFeO$_3$ & BiCoO$_3$ & BiNiO$_3$\\
\hline
space group &\textit{C2/c}           &\textit{C2/c}       &\textit{C2/c}       &\textit{R3c}       &\textit{P4mm}        &\textit{P\={1}}\\
 $a$, [\AA]   &9.8899(5)      &9.4641(4)  &9.5415(2)  &5.58102(4)& 3.71990(7) &5.3852(2)\\
 $b$, [\AA]  &5.8221(3)      &5.4790(2)  &5.61263(8) &5.58102(4)& 3.71990(7) &5.6498(2)\\
 $c$, [\AA]   &10.0462(5)     &9.5850(4)  &9.8632(2)  &13.8757(2)&4.71965(15)&7.7078(3)\\
$\alpha$, [$^\circ$]   &90         &90          &90          &90  &90 &91.9529(10)\\
$\beta$, [$^\circ$]    &108.300(3) & 108.568(3) &110.6584(12)&90  &90 &89.8097(9) \\
$\gamma$, [$^\circ$]   &90         &90          &90          &120 &90 &91.5411(9)\\
\hline
\end{tabular}
\end{table*}

\begin{table}[!htb]
\caption{Selected bond lengths in BiScO$_3$, BiCrO$_3$ and BiMnO$_3$, in \AA. $\Delta = \frac{1}{N} \sum_1^N(\frac{l_i-l_{av}}{l_{av}})^2$, and $l_{av} = \frac{1}{N} \sum_1^Nl_i$, where $l_i$ is the i-th bond.}
\label{sc_cr_mn_dist}
\begin{tabular}{|l|c|c|c|}
\hline
                &BiScO$_3$ &BiCrO$_3$&BiMnO$_3$\\
\hline
\textit{Me}1-O2x2         &2.08570 &1.98308 &1.90556\\
\textit{Me}1-O1x2         &2.10990 &1.99158 &2.19906\\
\textit{Me}1-O3x2         &2.15779 &1.97805 &1.98549\\

$\Delta$(\textit{Me}1-O)  &2.0$\times$10$^{-4}$&0.08$\times$10$^{-4}$&37.2$\times$10$^{-4}$\\
\hline
\textit{Me}2-O3x2         &2.09607 &1.97890 &1.94151\\
\textit{Me}2-O1x2         &2.11635 &1.99268 &1.92401\\
\textit{Me}2-O2x2         &2.13574 &2.01393 &2.24174\\
$\Delta$(\textit{Me}2-O)  &0.6$\times$10$^{-4}$&0.5$\times$10$^{-4}$&51.3$\times$10$^{-4}$\\
\hline
Bi-O3           &2.15352 &2.25697 &2.24562\\
Bi-O2           &2.19212 &2.23752 &2.21778\\
Bi-O1           &2.24633 &2.32584 &2.23928\\
Bi-O1           &2.55437 &2.44659 &2.46625\\
Bi-O3           &2.89682 &2.65246 &2.70996\\
Bi-O2           &3.01816 &2.78741 &2.83703\\
$\Delta$(Bi-O)  &18.7$\times$10$^{-3}$&7.0$\times$10$^{-3}$&9.9$\times$10$^{-3}$\\
\hline
\end{tabular}
\end{table}
\normalsize

\begin{table}[!htb]
\caption{Selected bond lengths in BiFeO$_3$ and BiCoO$_3$, in \AA.}
\label{fe_dist}
\begin{tabular}{|l|c|l|c|}
\hline
Fe-Ox3   &1.97139   &Co-O1    &1.71781\\
Fe-Ox3   &2.08310   &Co-O2x4  &2.01593\\
	
$\Delta$(\textit{Me}-O)  &8$\times$10$^{-4}$&&37$\times$10$^{-4}$\\
\hline
Bi-Ox3    &2.30512  &Bi-O2x4    &2.25155\\
Bi-Ox3    &2.51513  &Bi-O1x4    &2.79852\\

$\Delta$(Bi-O)  &1.9$\times$10$^{-3}$&&11.7$\times$10$^{-3}$\\

\hline
\end{tabular}
\end{table}

\begin{table*}[!htb]
\caption{Selected bond lengths in BiNiO$_3$, in \AA.}
\label{ni_dist}
\begin{tabular}{|l|c|l|c|l|c|l|c|}
\hline
Ni1-O2x2  &1.98294 &Ni2-O3x2 &2.04959 & Ni3-O5x2 &1.98075 & Ni4-O1x2 & 2.05408\\
Ni1-O3x2  &1.99200 &Ni2-O4x2 &2.07097 & Ni3-O2x2 &2.10662 & Ni4-O6x2 & 2.10927\\
Ni1-O4x2  &2.28697 &Ni2-O1x2 &2.13415 & Ni3-O6x2 &2.12071 & Ni4-O5x2 & 2.20769\\
$\Delta$(\textit{Me}-O)  &46$\times$10$^{-4}$&&3$\times$10$^{-4}$&&9$\times$10$^{-4}$&&9$\times$10$^{-4}$\\
\hline
Bi1-O3  &2.21311 &Bi2-O1 &2.03349&&&& \\
Bi1-O5  &2.26541 &Bi2-O4 &2.04058&&&&\\
Bi1-O2  &2.32491 &Bi2-O6 &2.09686&&&&\\
Bi1-O3  &2.41745 &Bi2-O2 &2.13787&&&&\\
Bi1-O1  &2.52033 &Bi2-O5 &2.24213&&&&\\
Bi1-O4  &2.67658 &Bi2-O6 &2.25756&&&&\\
$\Delta$(Bi-O)  &4.3$\times$10$^{-3}$&&1.7$\times$10$^{-3}$&&&&\\
\hline
\end{tabular}
\end{table*}

The bismuth manganite (BiMnO$_3$) has a highly distorted perovskite structure and has been regarded as a multiferroic material. The ferroelectricity has been analyzed within first-principles electronic structure calculations \cite{HillRabe}, and attributed to the chemical activity of the Bi \textit{6s}$^\mathit{2}$ lone pairs \cite{SeshadriHill}. 
BiMnO$_3$ is the only ferromagnet among the discussed Bi\textit{Me}O$_3$ compounds with a Curie temperature above 100 K. The largest saturation magnetization has been reported to be $3.92~\mu_B$ per formula unit~\cite{belik_mn_07}, which is close to the $4~\mu_B$ expected for the ferromagnetic state. The ferroelectric hysteresis loop has been observed in polycrystalline and thin film samples of BiMnO$_3$ \cite{SantosSSC}, although the measured ferroelectric polarization was much smaller (about $0.043$ $\mu$C/cm$^2$ at 200 K) than the one obtained in the first principle calculations for the experimental noncentrosymmetric structure (about 0.52 $\mu$C/cm$^2$) \cite{Shishidou_04}.

BiMnO$_3$ undergoes two phase transitions at temperatures 474K (monoclinic-monoclinic) and 770K (monoclinic-orthorhombic) \cite{atou_99, santos_02, montanari_05}. According to early experimental data BiMnO$_3$ was considered to have noncentrosymmetric \textit{C2} space group below 770 K \cite{atou_99, santos_02}. Recently the crystal structure of BiMnO$_3$ was re-examined by Belik \textit{et al.} \cite{belik_mn_07} and confirmed by neutron powder diffraction experiments by Montanari \textit{et al} \cite{montanari_07}. The new experiments reveal that BiMnO$_3$ below 770 K has a centrosymmetric \textit{C2/c} space group with parameters given in Table \ref{tab_struct}. A structural optimization performed using modern methods of electronic structure calculations has shown that the noncentrosymmetric \textit{C2} structure, which had been reported earlier converges to the new total energy minimum corresponding to the \textit{C2/c} structure with zero net polarization \cite{Shishidou, spaldin_07}. 

Since the \textit{C2/c} structure of BiMnO$_3$ has inversion symmetry the hypothesis that electric polarization arises due to bismuth lone pairs is no longer valid. The magnetic mechanism of inversion symmetry breaking was considered recently in work~\onlinecite{solovyev_08}. It was shown that the peculiar orbital ordering realized below 474 K gives rise to ferromagnetic (FM) interactions between nearest-neighbor spins which compete with longer-range antiferromagnetic (AFM) interactions. The solution of the low-energy model for \textit{3d} states in BiMnO$_3$ revealed that the true symmetry is expected to be \textit{Cc}. The solution of the realistic model indicates a noncollinear magnetic ground state, where the ferromagnetic order along one crystallographic axis coexists with the the hidden AFM order and a related ferroelectric polarization along two other axes~\cite{pzv_solovyev}.

The perovskite BiFeO$_3$ is ferroelectric with T$_c$ = 1103 K and antiferromagnetic with T$_N$ = 643 K and a canted spin structure~\cite{smolenskii_82}. The bulk single crystal has a rhombohedrally distorted perovskite structure with the space group \textit{R3c}~\cite{sosnowska_02} and lattice parameters presented in Table \ref{tab_struct}. The G-type collinear antiferromagnetic spin configuration has been modified by subjecting it to a long-range (620 \AA) modulation leading to a spiral modulated spin structure~\cite{sosnowska_82}. The spontaneous polarization of a single crystal is 3.5 $\mu$C/cm$^2$ along (001) direction and 6.1 $\mu$C/cm$^2$ in (111) direction at 77 K \cite{teague_70}. This value is significantly  smaller than spontaneous polarization of lead titanate (80 to 100 $\mu$C/cm$^2$ with the T$_C$ $\sim$763 K). However the heteroepitaxially constrained thin films of BiFeO$_3$ display a room-temperature spontaneous polarization of 50-60 $\mu$C/cm$^2$, an order of magnitude higher than in the bulk \cite{wang_03}. The spin polarized first-principles calculation within local spin density approximation (LSDA) using the pseudopotential VASP package with the optimizated lattice parameters for the bulk rhombohedral phase results in polarization of 6.61 $\mu$C/cm$^2$ \cite{wang_03} in excellent agreement with experiment. The thin films were shown to have tetragonal-like structure. For the \textit{P4mm} symmetry and lattice parameters measured for thin film the Barry-phase calculation yields a spontaneous polarization of 63.2 $\mu$C/cm$^2$ \cite{wang_03}, consistent with experimental data, but it was revealed that small changes of lattice parameters can lead to a dramatically different polarization. The magnetoelectric coefficient  ($dE$/$dH$, $E$ - electric field, $H$ - magnetic field) was measured to be 3 V/cm$\cdot$Oe at zero field \cite{wang_03}. Later on the values of remnant polarization were increased to 55 $\mu$C/cm$^2$ for (001) films, 80 $\mu$C/cm$^2$ for (101) films and about 100 $\mu$C/cm$^2$ for (111) films \cite{li_04}. The BiFeO$_3$ films grown on (111) have the rhombohedral structure as single crystals, whereas films grown on (101) or (001) are monoclinically distorted \cite{li_04}. The highest remnant polarization ever measured for a ferroelectric material, 146 $\mu$C/cm$^2$, has been reported for BiFeO$_3$ thin films with tetragonal crystal structure in work \cite{yun_04}. A wide range of measured polarization values were shown to be in consistent with the modern theory of polarization \cite{king_93, vanderbilt_93, resta_94}, which recognizes that polarization is a lattice of values rather than a single vector \cite{neaton_05}.  

BiCoO$_3$ is isotypic with BaTiO$_3$ and PbTiO$_3$ and has a tetragonal crystal structure with the lattice parameters $a$ = 3.71990(7) \AA, $b$ = 4.71965(15) \AA, $c/a$ = 1.269 at 5K \cite{belik_co_06}. BiCoO$_3$ is an insulator with T$_N$ = 470 K. It has C-type antiferromagnetic order where the magnetic moments of Co$^{3+}$ ions aligning antiferromagnetically in the $ab$ plane and antiferromagnetic y$ab$ layers stack ferromagnetically along  the $c$ axis \cite{belik_co_06}. 
The high-spin configuration of Co$^{3+}$ ions (S = 2) has been reported \cite{belik_co_06}. The magnetic moments are m = 3.24(2)$\mu_B$ for T = 5K and m = 2.93(2)$\mu_B$ at room temperature \cite{belik_co_06}. A reduction of the observed magnetic moment compared to the expected value 4$\mu_B$ may be ascribed to the covalency of Co-O bonds. The bond valence sums (BVS) at 300K obtained in \onlinecite{belik_co_06} were 3.14 for Bi and -2.13 for O2 corresponding  to the oxidation states +3 and -2, respectively. The BVS values were 2.68 for Co and -1.57 for O1 indicating the covalency effects. 
The spontaneous polarization in BiCoO$_3$ of 179 $\mu$C/cm$^2$ has been predicted from first-principle Berry-phase calculations \cite{uratani_05}. The experimental observation of a ferroelectric hysteresis loop is problematic since the resistivity appears too low for the large applied electric field. Therefore it was proposed that BiCoO$_3$ should be regarded as a pyroelectric rather than a ferroelectric material (in pyroelectrics the dipole moments can not be reoriented by external electric field) \cite{belik_co_06}. The C-type antiferromagnetic order with a reduced magnetic moment of 2.41 $\mu_B$ have  been also predicted from first-principle calculations \cite{uratani_05}. Spin-polarized calculations with C-type antiferromagnetic order for BiCoO$_3$ predicts an insulating ground state with an energy gap of 0.6 eV. 

BiNiO$_3$ has been found to be an insulating antiferromagnet (T$_N$ = 300 K) \cite{ishiwata_02} with a heavily distorted triclinic symmetry \textit{P\={1}}. The lattice constants are shown in Table~\ref{tab_struct}. X-ray powder diffraction data has revealed that Bi ions were disproportionately weighted towards Bi$^{3+}$ and Bi$^{5+}$ and therefore the oxidation state of the Ni ion was +2 rather than the expected +3 \cite{ishiwata_02, carlsson_08}. The Curie constant estimated from the magnetic susceptibility of BiNiO$_3$ is close to that expected for S = 1 rather than for S = 1/2 system. This fact as well as BVS confirms the divalent nature of Ni. The electronic structure of BiNiO$_3$ has been performed by the full-potential method within LDA+U ( \textit{U} = 8 eV,  \textit{J} = 0.95 eV) approximation with the G-type antiferromagnetic spin configuration \cite{azuma_07}. At ambient pressure a insulating solution with the charge-transfer gap of 0.8 eV was obtained in agreement with the value of 0.675 eV estimated from the electrical resistivity \cite{ishiwata_02}. From the powder neutron diffraction study it was found that the valence state of BiNiO$_3$ changes under pressure \cite{azuma_07}. Both neutron diffraction measurements and BVS show that under pressure the charge disproportionate state melts leading to the simultaneous charge transfer from Ni to Bi, so that the high-pressure phase is metallic Bi$^{3+}$Ni$^{3+}$O$_3$. This transition takes place at 4 GPa pressure and structure changes to the GdFeO$_3$-type with the \textit{Pbnm} symmetry \cite{azuma_07}. First-principle calculations also reproduce the metallic character of high-pressure phase.

\section{\label{result} Results and discussion}

\subsection{Results of LSDA calculation}
The electronic structure of Bi\textit{Me}O$_3$ (\textit{Me} = Sc, Cr, Mn, Fe, Co, Ni) series was calculated within the local spin density approximation (LSDA) \cite{lda, lda_1} with a linearized muffin-tin orbitals basis (LMTO)~\cite{lmto} using the STUTTGART TB-LMTO program (version LMTO47). For all compounds the experimental atomic positions and lattice constants shown in Table~\ref{tab_struct} were used. For the spin-polarized calculations we used the experimentally determined magnetic structures described in previous paragraphs (for BiFeO$_3$ the G-type AFM order was assumed). The resulting energy gaps (E$_g^{Calc.}$) as well as magnetic moments are shown in Table~\ref{lsda_results} along with the available experimental data. For BiScO$_3$, BiCrO$_3$, BiFeO$_3$ and BiCoO$_3$ LSDA gives an insulating solution. For BiMnO$_3$ LSDA results in a half-metallic solution although the material is known to be an insulator. The metallic state of BiNiO$_3$ obtained with LSDA calculations is also in contrast with experimental observations. The failure of LSDA is due to the improper treatment of Coulomb correlations. It is well known that LSDA+U  improves the description of correlation effects in transition metal oxides due to treating the strong Coulomb repulsion between localized $d$ states by adding a Hubbard-like term to the effective potential~\cite{lsda_u}. This calculation requires the Hubbard parameter $U$ and exchange interaction $J$. Although these parameters can be calculated by constrained LDA method~\cite{constr} at present work we just take typical values of $U$ = 3 eV, $J$ = 1 eV for BiMnO$_3$ and $U$ = 8 eV, $J$ = 1 eV for BiNiO$_3$. 

\begin{figure}[!ht]
\includegraphics[clip=true,angle=270,width=3in]{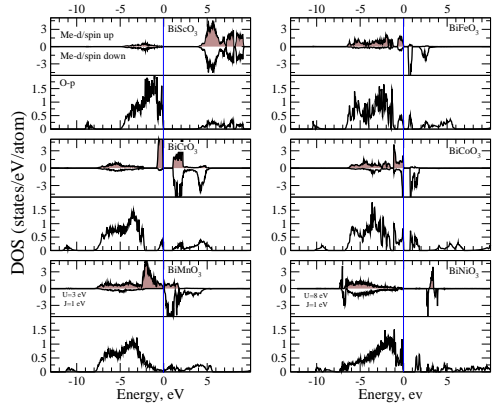}
\caption{\label{lsda} LSDA (LSDA+U with $U$ = 3 eV, $J$ = 1 eV for BiMnO$_3$ and $U$ = 8 eV, $J$ = 1 eV for BiNiO$_3$) partial \textit{Me} \textit{3d} and O \textit{2p} density of states for Bi\textit{Me}O$_3$ (\textit{Me} = Sc, Cr, Mn, Fe, Co, Ni) compounds. The Fermi level corresponds to zero. (Colour in on-line version.)}
\end{figure}

The total and partial densities of states (DOS) obtained for Bi\textit{Me}O$_3$ series within LSDA and LSDA+U (for BiNiO$_3$ and BiMnO$_3$) are presented in Figures~\ref{lsda} and ~\ref{lsda_pdos}. One can see that the valence band for all Bi\textit{Me}O$_3$ compounds is formed by the \textit{Me} \textit{3d} states hybridized with the O \textit{2p} states. The low lying states at energy about -10 eV comes from Bi \textit{6s} states, the so-called ``lone pair''. These states are only slightly hybridized with \textit{2p} states of oxygen. 

\begin{table*}[htb]
\caption{ The calculated (E$_g^{Calc.}$) and estimated from XES and XAS spectra ($\Delta_g$) values of energy gap in comparison with the experimental data taken from literature (E$_g^{Lit.}$). The calculated magnetic moments on the \textit{Me} ion (m$_{Me}$) along with experimental estimations (m$_{Me}^{Lit.}$) for Bi\textit{Me}O$_3$ compounds. The different values of magnetic moments in case of \textit{Me} = Cr, Mn, Ni correspond to the nonequivalent \textit{Me} atoms in the unit cell.}
\label{lsda_results}
\begin{tabular}{|l|c|c|c|c|c|c|}
\hline
Compound &E$_g^{Calc.}$ [eV]&$\Delta_g$ [eV]& E$_{g}^{Lit.}$ [eV] & m$_{Me}$ [$\mu_B$]& m$_{Me}^{Lit.}$ [$\mu_B$]&Config. \textit{Me}$^{3+}$ \\
\hline
BiScO$3$ &3.3 & 2.6&- & 0 & -& $d^0$\\
BiCrO$_3$&0.88& 1.4&- &2.63, 2.65& 2.55~\cite{belik_08}& $d^3$\\
BiMnO$_3$&0.33& 0.9&insulator~\cite{tokura_03}& 3.65, 3.64&3.2~\cite{moreira_02}& $d^4$\\
BiFeO$_3$&0.51 & 0.9&1.3~\cite{higuchi_08}, 2.5~\cite{gao_06}&3.54 & 3.75~\cite{sosnowska_02}& $d^5$\\
BiCoO$_3$&0.72 & 1.7& insulator~\cite{belik_co_06}&2.41&3.24\cite{belik_co_06}& $d^6$\\
BiNiO$_3$&1.23 & 1.1& 0.675~\cite{ishiwata_02}& 1.7; 1.67 & 1.76~\cite{carlsson_08}& $d^7$\\
\hline
\end{tabular}
\end{table*}

BiScO$_3$, BiCrO$_3$, and  BiMnO$_3$ all have a monoclinic structure and can be compared directly. The bondlengths for these three compounds are summarized in Table~\ref{sc_cr_mn_dist}. 
Although the space groups of these compounds are the same, the \textit{Me}O$_6$ octahedra distortions are different for each and most pronounced in the case of BiMnO$_3$. The parameter $\Delta$ in Table~\ref{sc_cr_mn_dist} indicates the degree of \textit{Me}O$_6$ octahedra distortion. The largest $\Delta$ corresponds to BiMnO$_3$ while in BiCrO$_3$ the CrO$_6$ octahedron is almost undistorted. The shortest Bi---O bond lengths also shrink from Sc to Mn. The oxygen atoms surrounding Bi do not form regular polyhedra so we take the six nearest neighbors and estimate the variations between the compounds. All three compounds display rather strong deviation from average Bi---O bond but the largest one is found for the BiScO$_3$ system. BiFeO$_3$ and BiCoO$_3$ have different structures. Five oxygen atoms surrounding the Co ion form a pyramid (see Table~\ref{fe_dist}). This pyramid is contracted so the distortion parameter is rather large. In BiFeO$_3$ the FeO$_6$ octahedron is less distorted than in BiMnO$_3$ system. The oxygen atoms are distributed around Bi in BiFeO$_3$ almost evenly while in BiCoO$_3$ rather irregular arrangement of oxygen atoms can be seen from Bi---O bondlengths shown in Table~\ref{fe_dist}. In Table~\ref{ni_dist} the distances between Bi---O and Ni---O for BiNiO$_3$ are shown. The low symmetry of BiNiO$_3$ leads to the four nonequivalent Ni atoms and two nonequivalent Bi atoms in the unit cell. Among four Ni octahedra formed by oxygen atoms the one around Ni1 is the most distorted. The oxygen neighborhood around the Bi atoms is rather uniform. 

For BiScO$_3$ and BiCrO$_3$ an insulating solution was obtained with LSDA calculations (see Figure~\ref{lsda}). BiScO$_3$ has a formal electronic configuration $d^0$ and calculated energy gap of 3.3 eV. The valence band is formed by O \textit{2p} states while the conduction band is composed of Sc \textit{3d} states hybridized with O \textit{2p} states. BiCrO$_3$ has three electrons occupying $t_{2g}$ orbitals (see Figure~\ref{lsda}). The energy gap of 0.88 eV and magnetic moment of 2.63 and 2.65 $\mu_B$ (for two nonequivalent Cr atoms) was obtained in spin-polarized calculation in good agreement with the value 2.55 $\mu_B$ obtained in neutron diffraction experiments~\cite{belik_08}. Both the valence and conduction bands are formed mainly by Cr \textit{3d} states with small admixture of O \textit{2p} states. LSDA+U produces the insulating state for ferromagnetically ordered BiMnO$_3$ in agreement with experiment (see Figure~\ref{lsda}). The calculated magnetic moments for two nonequivalent Mn ions are 3.64 and 3.65 $\mu_B$ which is in agreement with experimental values shown in Table~\ref{lsda_results}. Both the O \textit{2p} and the Mn \textit{3d} bandwidths are wider than those of BiCrO$_3$ but the main contribution to the valence band comes from Mn \textit{3d} states. There is also a strong overlap of O \textit{2p} and Mn \textit{3d} bands. 

LSDA calculations result in an insulating solution for BiFeO$_3$ with the energy gap of 0.51 eV. This is much smaller than experimental estimates~\cite{higuchi_08, gao_06}. However the magnetic moment value of 3.54 $\mu_B$ is in good agreement with experiment (see Table ~\ref{lsda_results}). The calculated DOS agrees well with the results of previous \textit{ab initio} calculations \cite{neaton_05}. Within LSDA the high spin (HS) state for the $d^5$ configuration of Fe ion was obtained as shown in  Figure~\ref{lsda}. Both the valence and conduction bands are formed predominantly by Fe \textit{3d} states hybridized with O \textit{2p} states. 

The LSDA calculations for BiCoO$_3$ suggest it is an insulator with an energy gap of 0.72 eV. We obtained HS state of $d^6$ electrons with magnetic moment of 2.41$\mu_B$ which is smaller than 4 $\mu_B$ (expected for HS state) due to covalency effects.

With the LSDA+U method BiNiO$_3$ is insulator with an energy gap of 1.23 eV. This value is twice as large as the experimental result from Reference \onlinecite{ishiwata_02} but in good agreement with the present experimental estimation (see Table \ref{lsda_results}). The values of the calculated magnetic moments given in Table~\ref{lsda_results} are in agreement with the experimental one. From the density of states it can be seen that the top of the valence band and the bottom of the conduction band are formed by Bi1 \textit{6s} and Bi2 \textit{6s} states hybridized with O \textit{2p} states. The occupancies calculated for the \textit{d} states of the four non-equivalent Ni atoms in the LSDA+U calculation are 8.2, 8.36, 8.23, and 8.49 indicating that Ni has a 2+ valence which is in agreement with calculations performed in the literature \cite{azuma_07}.

\subsection{Experimental results and discussion}

\begin{figure}[htb]
\includegraphics[clip=true,angle=270,width=3in]{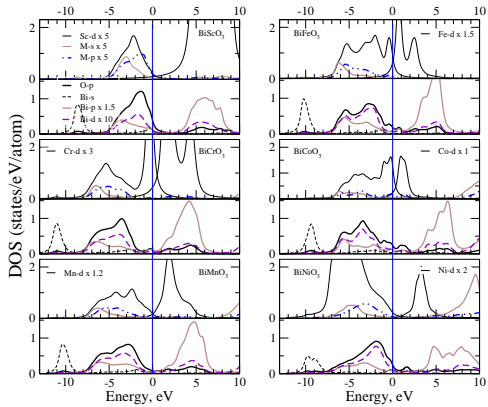}
\caption{\label{lsda_pdos} Calculated LSDA (LSDA+U for BiMnO$_3$ and BiNiO$_3$) partial density of states for Bi\textit{Me}O$_3$ (\textit{Me} = Sc, Cr, Mn, Fe, Co, Ni) compounds. The DOS have been convoluted with the Fermi step at T = 300 K and broadened by 0.3 eV using a Lorentzian function to mimic the experimental resolution. The Fermi level corresponds to zero. (Colour in on-line version.)}
\end{figure}

\begin{figure}[!ht]
\includegraphics[clip=true,width=3in]{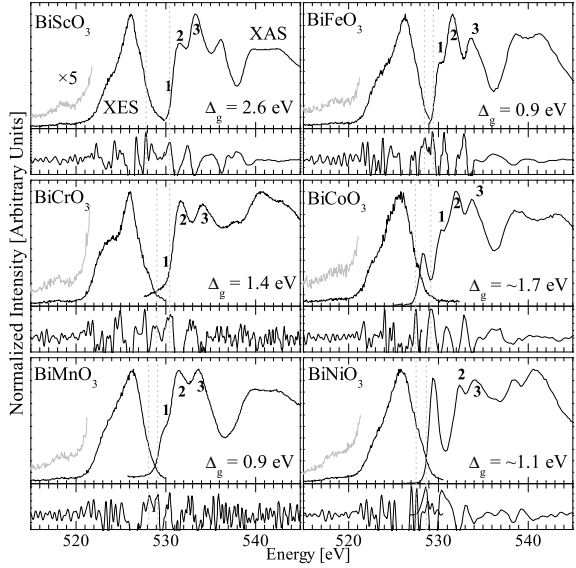}
\caption{\label{O_emission} Measured oxygen K$_{\alpha}$ X-ray emission spectra (XES) and O \textit{1s} X-ray absorption spectra (XAS) of Bi\textit{Me}O$_3$ compounds. The XAS is measured in total fluorescent yield (TFY) mode. The energy gap is estimated using the peaks in the second derivative, plotted below the corresponding spectra. The Bi \textit{6s} --- O \textit{2p} hybridization peak at about 519 eV has been magnified by a factor of 5.}
\end{figure}

The O K$_{\alpha}$ XES and O \textit{1s} XAS measurements of Bi\textit{Me}O$_3$ (\textit{Me} = Sc, Cr, Mn, Fe, Co, Ni), which probe the occupied and vacant O \textit{2p} states, respectively, are presented in Figure~\ref{O_emission}. The fine structure and energy distribution of the O K$_{\alpha}$ XES matches the O \textit{2p} occupied DOS obtained from LSDA calculations (see Figure~ \ref{lsda_pdos}). The band gap for these materials was estimated using the peaks in the second derivative, this method has been shown to work well with O K$_{\alpha}$ XES and \textit{1s} XAS \cite{kurmaev_08}. A small amount of hybridization between the Bi \textit{6s}- and O \textit{2p}-states is visible at about 519 eV in all compounds, as suggested by the calculated O \textit{2p}-states.

\begin{figure}[!ht]
\includegraphics[clip=true,width=3in]{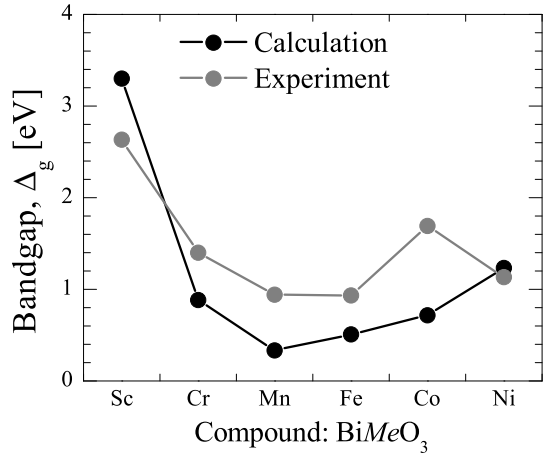}
\caption{\label{energy gap} The experimental (gray dots) and calculated (black dots) energy gaps for the Be\textit{Me}O$_3$ compounds.}
\end{figure}

The estimated energy gaps from the experimental spectra and the calculations are not exactly the same. This is expected, since LSDA calculations are known to provide a underestimation of the energy gap, and the LSDA+U calculations used a typical value for \textit{U} rather than a material-specific value. We have found that the calculated energy gaps are mostly within 0.5 eV of the experimental energy gaps and, more importantly, the trend in the calculated energy gaps is essentially the same as that in the experimental energy gaps for these materials. The energy gaps are shown in Figure~\ref{energy gap}, note the similar shape in the experimental and calculated curves. 

\begin{figure}[!ht]
\includegraphics[clip=true,width=3in]{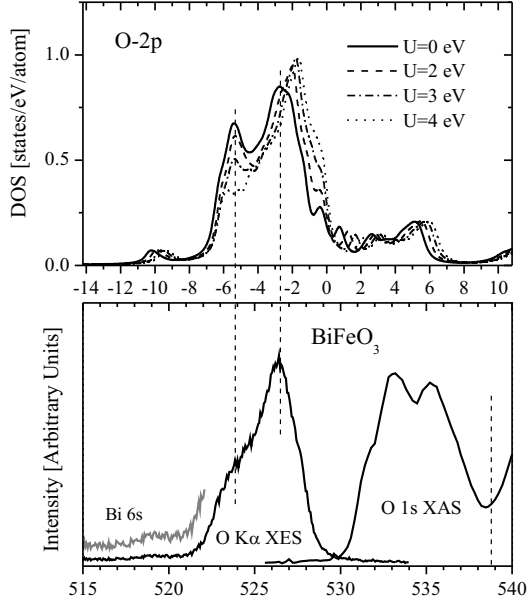}
\caption{\label{fe_o2p} The comparison between the O K$_\alpha$ XES and O \textit{1s} XAS (lower panel) and the LSDA+U O \textit{2p} DOS (upper panel) of BiFeO$_3$. In the upper panel, the calculated LSDA+U O \textit{2p} DOS for several different values of U are shown. The O \textit{2p} DOSes were broadened by 0.3 eV using a Lorentzian function to mimic the experimental resolution. The Fermi level corresponds to zero.}
\end{figure}

\begin{figure}[!ht]
\includegraphics[clip=true,width=3in]{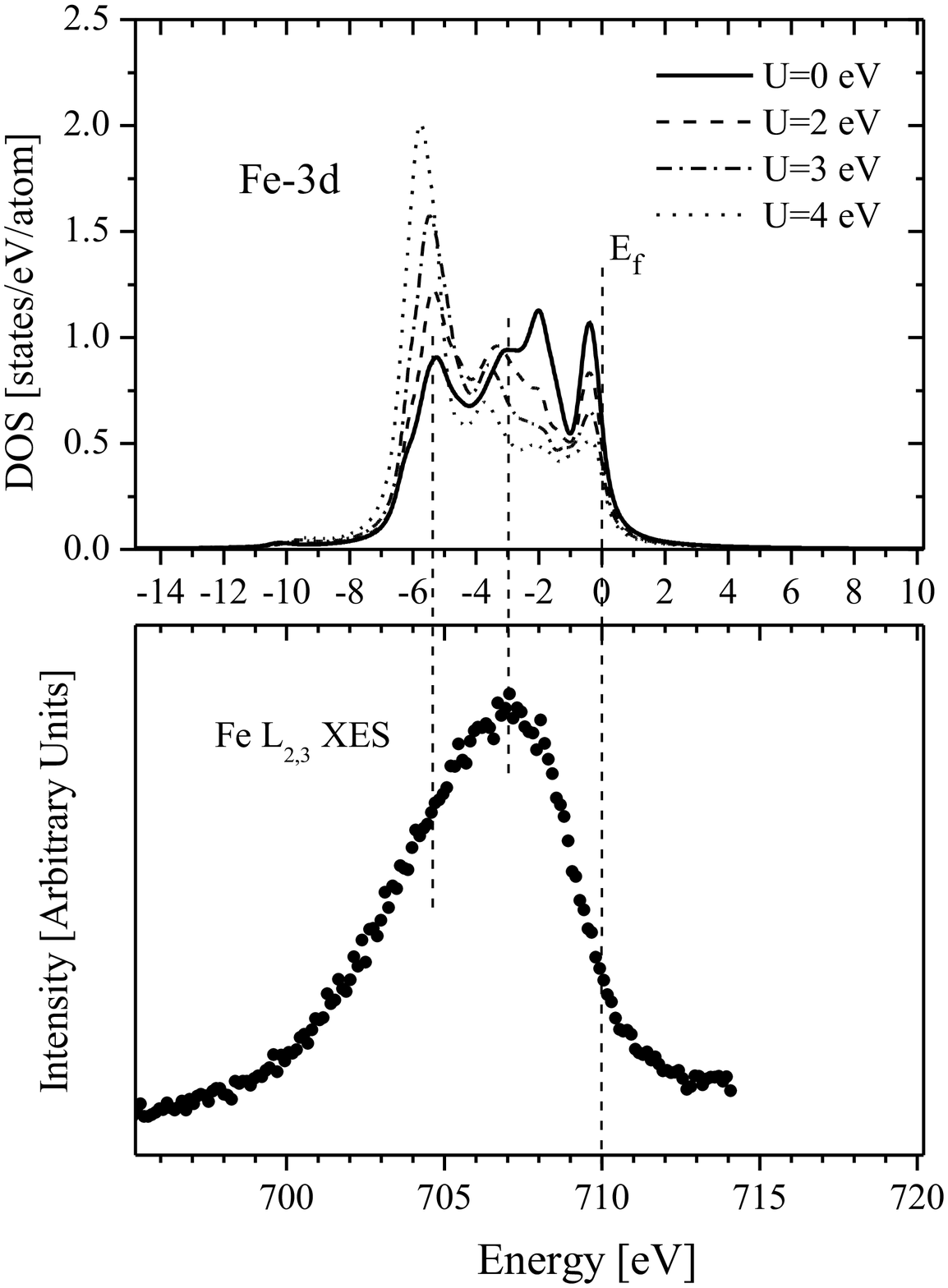}
\caption{\label{fe_fe3d} The comparison between the Fe \textit{L}$_\mathit{2,3}$ XES (lower panel) and the LSDA+U Fe \textit{3d} DOSes (upper panel) of BiFeO$_3$. In the upper panel, the calculated LSDA+U Fe \textit{L}$_\mathit{2,3}$ DOS for several different values of \textit{U} are shown. The Fe $3d$ DOS has been convoluted with the Fermi step at T = 300 K and broadend by 0.3 eV using a Lorentzian to mimic the experimental resolution. The Fermi level corresponds to zero.}
\end{figure}

The comparison of the oxygen and iron X-ray spectra of BiFeO$_3$ with the calculated O \textit{2p} and Fe \textit{3d} DOSes is shown in Figure~\ref{fe_o2p} and Figure~\ref{fe_fe3d}, respectively (solid line). The Fermi level position for the Fe \textit{L}$_\mathit{3}$ XES was determined using the XPS Fe \textit{2p}$_\mathit{3/2}$ binding energy (710 eV) \cite{kharel_09}.  In the absence of XPS O \textit{1s} data the O K$_{\alpha}$ XES is compared with O \textit{2p} DOS by alignment of Bi \textit{6s} related subbands.
The LSDA calculation reproduces the energy position of the Fe \textit{L}$_\mathit{3}$ XES (which probes occupied Fe \textit{3d} states) with respect to the Fermi level. On the other hand, the available LSDA+U calculations \cite{neaton_05, higuchi_08} show a low-energy shift of Fe \textit{3d} band which contradicts the experimental Fe \textit{L}$_\mathit{3}$ XES. The spectral weight of Fe \textit{3d}- and O \textit{2p} states is redistributed with increasing U value within LSDA+U calculations, as shown in Figures~\ref{fe_o2p} and \ref{fe_fe3d}. As the \textit{U} value rises the Fe \textit{3d} states are shifted to lower energies (Figure~\ref{fe_fe3d}). At the same time the O \textit{2p} states (Figure~\ref{fe_o2p}) are shifted towards Fermi level. Taking this trend into account  both the occupied O \textit{2p}- and Fe \textit{3d} states obtained with \textit{U} = 0 eV are in better agreement with experimental spectra than the one calculated with a nonzero \textit{U}. The Fe \textit{L}$_\mathit{2,3}$ XES spectrum has some intensity at and just below the Fermi level while the O K$_{\alpha}$ XES spectrum indicates a clear energy gap (see the lower panel of  Figure~\ref{fe_o2p}) in this energy region. In a recently published paper \cite{higuchi_08} it is concluded that LSDA+U calculations (\textit{U} = 2 eV) are a better match to experimental Fe \textit{L}$_\mathit{3}$ XES of BiFeO$_3$ than a LSDA (where \textit{U} = 0 eV) calculation. However, the comparison procedures used in matching calculations to experimental spectra are not described in Reference \onlinecite{higuchi_08}. Based on the comparison from Reference ~\onlinecite{higuchi_08} one can estimate an XPS Fe \textit{2p}$_\mathit{3/2}$ binding energy of 713 eV which contradicts the available experimental data (710 eV) \cite{kharel_09}. 

As we have shown in Reference~\onlinecite{skorikov}, the O \textit{2p} DOS for binary oxides of non-transition elements occupy a whole energy range of excited $d$ states of cations, involved in formation of the chemical bonding in solids. In Bi\textit{Me}O$_3$ compounds, as seen in Figure~\ref{lsda_pdos}, a correspondence of energies in the main maxima of O \textit{2p} and Bi \textit{6d} DOS occurs. The main maxima of the calculated \textit{Me} \textit{3d} DOS are also close to these energies. The non-bonding localized \textit{3d} states weakly mix with O \textit{2p} states near Fermi level.

All O K$_{\alpha}$ XES consist of low-energy subband located at $\sim$2.5 eV with respect to the main maximum. As it was shown in Reference~\onlinecite{skorikov}, the low-energy subband of O K$_{\alpha}$ XES of all binary oxides is directly connected with the $s$ valence states of the cation. The energy separation of this subband from the main maximum depends in particular on the period number of \textit{Me}. For ZnO it is $\sim$ 2.4 eV, which coincides with the separation of subband from the main band of O K$_{\alpha}$ XES for Bi\textit{Me}O$_3$ compounds. The Bi \textit{6s} states are located $\sim$ 5 eV below the \textit{Me} \textit{4s} states, this is why they hybridize only weakly with the O \textit{2p} states (see Figure~\ref{lsda_pdos}).

The near edge fine structure of the O \textit{1s} XAS (see Figure~\ref{O_emission}) consists of three peaks (labeled as 1, 2 and 3) separated by $\sim$ 1.5 - 2 eV for all compounds. 
There is a similar peak structure in the calculated unoccupied O \textit{2p} DOS (Figure~ \ref{lsda_pdos}) in spite of the fact that the band gap (E$_g^{Calc.}$) is not reproduced in LSDA calculations. Indeed, the lack of the appropriate energy gap does not effect the correct reproduction of the energy distance between the main maximum of the O K$_{\alpha}$ XES and the first peak of the O \textit{1s} XAS ($\sim$ 5 eV) in the LSDA+U calculations. This further demonstrates the well-known fact that O \textit{2p} states are less influenced by correlation effects than $d$ states \cite{kurmaev_08}. 

The O \textit{1s} XAS spectra of BiNiO$_3$ and BiCoO$_3$ show additional peaks near the Fermi level similar to those which were observed in the O \textit{1s} XAS of Li$_x$Ni$_{1-x}$O \cite{kuiper_89}, LiCoO$_2$ \cite{galakhov_02} and LaNiO$_3$ \cite{abbate_02}. These additional peaks were attributed in References \onlinecite{kuiper_89, abbate_02} to the formation of O \textit{2p} hole electronic states because of the energy disadvantage associated with a Ni$^{3+}$ charge state. The Co$^{3+}$ charge state is more stable, hence the intensity of the hole-like peak in BiCoO$_3$ is much lower than that in BiNiO$_3$. Indeed, the ``hole effect'' in BiNiO$_3$ is even larger than that in Li$_x$Ni$_{1-x}$O \cite{kuiper_89}. Among the materials studied in Reference~\onlinecite{kuiper_89} the intensity of the O \textit{2p} hole peak was found to be largest for Li$_{0.5}$Ni$_{0.5}$O where the hole peak had $\sim$70\% of the intensity of the main maximum of the O \textit{1s} XAS. Similar intensity ratios have been reported for LaNiO$_3$ \cite{abbate_02}. In BiNiO$_3$, however, the relative intensity of the O \textit{2p} hole peak is about 100\% of the intensity of the main O \textit{2p} band. The O \textit{2p} holes are created in order to maintain the neutrality of the compound with the Ni$^{2+}$ valence state. They are clearly observed in the O \textit{1s} XAS for Ni oxides LaNiO$_3$ and BiNiO$_3$. In the LSDA+U calculation the gap arises betwen the states predominantly formed by Bi \textit{6s} states strongly hybridized with O \textit{2p} states. The Ni atoms have a 2+ valency in agreement with experimental findings.

Now let us discuss the origin of ferroelectricity in BiMnO$_3$ and BiFeO$_3$. From Figure~\ref{lsda_pdos} it is clear that the Bi \textit{6s} states in all compounds are not hybridized with the Bi \textit{6p} states. As was supposed in Reference~\onlinecite{brink_08}, these Bi \textit{6s} lone pairs are slightly hybridized with O \textit{2p} states, and can induce dipole moment. However, we could not find any difference in degree of Bi \textit{6s}-O \textit{2p} hybridization for BiMnO$_3$ and BiFeO$_3$ compared to that of other Bi\textit{Me}O$_3$ compounds where no ferroelectricity is observed. The stereochemical activity of the Bi lone pair induces the electric moment but if the symmetry of the system includes inversion the net polarization turns to zero. The space group for the monoclinic structure of BiMnO$_3$ contains inversion while the rhombohedral BiFeO$_3$ has \textit{R3c} symmetry without inversion. Therefore the origin of ferroelectricity in these two compounds seems to be different. In BiMnO$_3$ the emergence of electric moment is connected with inversion symmetry breaking by the AFM order induced by particular orbital order below T = 474K~\cite{solovyev_08, pzv_solovyev}. Another scenario for the onset of electric moment is possible in BiFeO$_3$: an extensive investigation of ferroelectricity in BiFeO$_3$ in the framework of density functional theory was reported in Reference~\onlinecite{neaton_05}. It was mentioned that the driving force of the ferroelectric distortion is the Bi lone pair. The calculated values of polarization agree well with experimental results~\cite{wang_03}. 

Among the Bi\textit{Me}O$_3$ series there are three compounds with no inversion symmetry: BiMnO$_3$ (monoclinic structure, but inversion is broken by AFM order), BiFeO$_3$ (rhombohedral structure) and BiCoO$_3$ (tetragonal structure). The ferroelectric properties of the BiMnO$_3$ and BiFeO$_3$ are confirmed experimentally while a ferroelectric hysteresis loop has not yet been observed for BiCoO$_3$. One possible reason could be a low resistivity to the applied electric field as was pointed out in Reference~\onlinecite{belik_co_06}. The crystal structures of the BiScO$_3$ (\textit{C2/c}), BiCrO$_3$ (\textit{C2/c}), and BiNiO$_3$ (\textit{P\={1}}) compounds include the inversion operation. 

\section{\label{sum} Summary}
In conclusion, the electronic structure of the Bi\textit{Me}O$_3$ (\textit{Me} = Sc, Cr, Mn, Fe, Co, Ni) series was studied by soft X-ray emission and absorption spectroscopy. Experimental spectra were found to be in good agreement with spin-polarized electronic structure calculations. The presence of holes in the O \textit{1s} XAS spectrum of BiNiO$_3$ was attributed to the $2+$ valency of Ni. For all Bi\textit{Me}O$_3$ (\textit{Me} = Sc, Cr, Mn, Fe, Co, Ni) compounds the band gap values were estimated from O K$_{\alpha}$ XES and O \textit{1s} XAS spectra. In case of BiFeO$_3$ this estimation results in the band gap value $\sim$ 0.9 eV. The X-ray spectra for multiferroic and nonmultiferroic compounds in the series do not reveal any difference in degree of Bi \textit{6s}-O \textit{2p} hybridization. It was concluded that for nonzero electric polarization the stereochemical activity of the Bi lone pair should be accompanied with the inversion symmetry breaking. This condition is satisfied in case of BiMnO$_3$, BiFeO$_3$ and BiCoO$_3$ but the electric polarization arises due to the different reasons. In BiMnO$_3$ noncolinear magnetic order with AFM component breaks the inversion symmetry and allow net polarization. In BiFeO$_3$ the rhombohedral crystal structure does not contain inversion and the crucial role is played by Bi lone pairs. BiCoO$_3$ is supposed to be ferroelectric but a ferroelectric hysteresis loop has not been observed as of now so apparently BiCoO$_3$ should be considered as a pyroelectric material. 

\begin{acknowledgments}
We acknowledge support of the Research Council of the President of the Russian Federation (Grants NSH-1929.2008.2,  NSH-1941.2008.2 and MK-3227.2008.2(PZV)), the Russian Science Foundation for Basic Research (Project 08-02-00148), the  Dynasty Foundation (PZV), the Natural Sciences and Engineering Research Council of Canada (NSERC), and the Canada Research Chair program.
\end{acknowledgments}

\bibliography{multiferroics}

\begin{thebibliography}{60}
\expandafter\ifx\csname natexlab\endcsname\relax\def\natexlab#1{#1}\fi
\expandafter\ifx\csname bibnamefont\endcsname\relax
  \def\bibnamefont#1{#1}\fi
\expandafter\ifx\csname bibfnamefont\endcsname\relax
  \def\bibfnamefont#1{#1}\fi
\expandafter\ifx\csname citenamefont\endcsname\relax
  \def\citenamefont#1{#1}\fi
\expandafter\ifx\csname url\endcsname\relax
  \def\url#1{\texttt{#1}}\fi
\expandafter\ifx\csname urlprefix\endcsname\relax\def\urlprefix{URL }\fi
\providecommand{\bibinfo}[2]{#2}
\providecommand{\eprint}[2][]{\url{#2}}

\bibitem[{\citenamefont{Smolenskii et~al.}(1961)\citenamefont{Smolenskii,
  Isupov, Agranovskaya, and Kranik}}]{smolenskii_61}
\bibinfo{author}{\bibfnamefont{G.}~\bibnamefont{Smolenskii}},
  \bibinfo{author}{\bibfnamefont{V.}~\bibnamefont{Isupov}},
  \bibinfo{author}{\bibfnamefont{A.}~\bibnamefont{Agranovskaya}},
  \bibnamefont{and} \bibinfo{author}{\bibfnamefont{N.}~\bibnamefont{Kranik}},
  \bibinfo{journal}{Sov. Phys. Solid State} \textbf{\bibinfo{volume}{2}},
  \bibinfo{pages}{2651} (\bibinfo{year}{1961}).

\bibitem[{\citenamefont{Smolenskii et~al.}(1963)\citenamefont{Smolenskii,
  Yudin, Sher, and Stolypin}}]{smolenskii_63}
\bibinfo{author}{\bibfnamefont{G.}~\bibnamefont{Smolenskii}},
  \bibinfo{author}{\bibfnamefont{V.}~\bibnamefont{Yudin}},
  \bibinfo{author}{\bibfnamefont{E.}~\bibnamefont{Sher}}, \bibnamefont{and}
  \bibinfo{author}{\bibfnamefont{Y.~E.} \bibnamefont{Stolypin}},
  \bibinfo{journal}{Sov. Phys. JETP} \textbf{\bibinfo{volume}{16}},
  \bibinfo{pages}{622} (\bibinfo{year}{1963}).

\bibitem[{\citenamefont{Scott}(2007)}]{scott_07}
\bibinfo{author}{\bibfnamefont{J.~F.} \bibnamefont{Scott}},
  \bibinfo{journal}{Nature Materials} \textbf{\bibinfo{volume}{6}},
  \bibinfo{pages}{256} (\bibinfo{year}{2007}).

\bibitem[{\citenamefont{Spaldin and Fiebig}(2005)}]{spaldin_05}
\bibinfo{author}{\bibfnamefont{N.~A.} \bibnamefont{Spaldin}} \bibnamefont{and}
  \bibinfo{author}{\bibfnamefont{M.}~\bibnamefont{Fiebig}},
  \bibinfo{journal}{Science} \textbf{\bibinfo{volume}{309}},
  \bibinfo{pages}{391} (\bibinfo{year}{2005}).

\bibitem[{\citenamefont{Ederer and Spaldin}(2005)}]{ederer_05}
\bibinfo{author}{\bibfnamefont{C.}~\bibnamefont{Ederer}} \bibnamefont{and}
  \bibinfo{author}{\bibfnamefont{N.~A.} \bibnamefont{Spaldin}},
  \bibinfo{journal}{Current Opinion in Solid State and Materials Science}
  \textbf{\bibinfo{volume}{9}}, \bibinfo{pages}{128} (\bibinfo{year}{2005}).

\bibitem[{\citenamefont{Higuchi et~al.}(2008)\citenamefont{Higuchi, Liu, Yao,
  Glans, Guo, C.~Chang, Sakamoto, Itoh, Shimura, Yogo et~al.}}]{higuchi_08}
\bibinfo{author}{\bibfnamefont{T.}~\bibnamefont{Higuchi}},
  \bibinfo{author}{\bibfnamefont{Y.-S.} \bibnamefont{Liu}},
  \bibinfo{author}{\bibfnamefont{P.}~\bibnamefont{Yao}},
  \bibinfo{author}{\bibfnamefont{P.-A.} \bibnamefont{Glans}},
  \bibinfo{author}{\bibfnamefont{J.}~\bibnamefont{Guo}},
  \bibinfo{author}{\bibfnamefont{Z.~W.} \bibnamefont{C.~Chang}},
  \bibinfo{author}{\bibfnamefont{W.}~\bibnamefont{Sakamoto}},
  \bibinfo{author}{\bibfnamefont{N.}~\bibnamefont{Itoh}},
  \bibinfo{author}{\bibfnamefont{T.}~\bibnamefont{Shimura}},
  \bibinfo{author}{\bibfnamefont{T.}~\bibnamefont{Yogo}}, \bibnamefont{et~al.},
  \bibinfo{journal}{Phys. Rev. B} \textbf{\bibinfo{volume}{78}},
  \bibinfo{pages}{085106} (\bibinfo{year}{2008}).

\bibitem[{\citenamefont{Kang et~al.}(2005)\citenamefont{Kang, Han, Park, Wi,
  Lee, Kim, Song, Shin, Jo, and Min}}]{kang_05}
\bibinfo{author}{\bibfnamefont{J.-S.} \bibnamefont{Kang}},
  \bibinfo{author}{\bibfnamefont{S.~W.} \bibnamefont{Han}},
  \bibinfo{author}{\bibfnamefont{J.-G.} \bibnamefont{Park}},
  \bibinfo{author}{\bibfnamefont{S.~C.} \bibnamefont{Wi}},
  \bibinfo{author}{\bibfnamefont{S.~S.} \bibnamefont{Lee}},
  \bibinfo{author}{\bibfnamefont{G.}~\bibnamefont{Kim}},
  \bibinfo{author}{\bibfnamefont{H.~J.} \bibnamefont{Song}},
  \bibinfo{author}{\bibfnamefont{H.}~\bibnamefont{Shin}},
  \bibinfo{author}{\bibfnamefont{W.}~\bibnamefont{Jo}}, \bibnamefont{and}
  \bibinfo{author}{\bibfnamefont{B.}~\bibnamefont{Min}},
  \bibinfo{journal}{Phys. Rev. B} \textbf{\bibinfo{volume}{71}},
  \bibinfo{pages}{092405} (\bibinfo{year}{2005}).

\bibitem[{\citenamefont{Belik et~al.}(2008)\citenamefont{Belik, Iikubo, Kodama,
  Igawa, Shamoto, and Takayama-Muromachi}}]{belik_08}
\bibinfo{author}{\bibfnamefont{A.~A.} \bibnamefont{Belik}},
  \bibinfo{author}{\bibfnamefont{S.}~\bibnamefont{Iikubo}},
  \bibinfo{author}{\bibfnamefont{K.}~\bibnamefont{Kodama}},
  \bibinfo{author}{\bibfnamefont{N.}~\bibnamefont{Igawa}},
  \bibinfo{author}{\bibfnamefont{S.}~\bibnamefont{Shamoto}}, \bibnamefont{and}
  \bibinfo{author}{\bibfnamefont{E.}~\bibnamefont{Takayama-Muromachi}},
  \bibinfo{journal}{Chem. Mater.} \textbf{\bibinfo{volume}{20}},
  \bibinfo{pages}{3765} (\bibinfo{year}{2008}).

\bibitem[{\citenamefont{Belik et~al.}(2007{\natexlab{a}})\citenamefont{Belik,
  Iikubo, Yokosawa, Kodama, Igawa, Shamoto, Azuma, Takano, Kimoto, Matsui
  et~al.}}]{belik_mn_07}
\bibinfo{author}{\bibfnamefont{A.~A.} \bibnamefont{Belik}},
  \bibinfo{author}{\bibfnamefont{S.}~\bibnamefont{Iikubo}},
  \bibinfo{author}{\bibfnamefont{T.}~\bibnamefont{Yokosawa}},
  \bibinfo{author}{\bibfnamefont{K.}~\bibnamefont{Kodama}},
  \bibinfo{author}{\bibfnamefont{N.}~\bibnamefont{Igawa}},
  \bibinfo{author}{\bibfnamefont{S.}~\bibnamefont{Shamoto}},
  \bibinfo{author}{\bibfnamefont{M.}~\bibnamefont{Azuma}},
  \bibinfo{author}{\bibfnamefont{M.}~\bibnamefont{Takano}},
  \bibinfo{author}{\bibfnamefont{K.}~\bibnamefont{Kimoto}},
  \bibinfo{author}{\bibfnamefont{Y.}~\bibnamefont{Matsui}},
  \bibnamefont{et~al.}, \bibinfo{journal}{J. Am. Chem. Soc.}
  \textbf{\bibinfo{volume}{129}}, \bibinfo{pages}{971}
  (\bibinfo{year}{2007}{\natexlab{a}}).

\bibitem[{\citenamefont{Belik et~al.}(2006{\natexlab{a}})\citenamefont{Belik,
  Iikubo, Kodama, Igawa, Shamoto, Maie, Nagai, Matsui, Stefanovich, Lazoryak
  et~al.}}]{belik_06}
\bibinfo{author}{\bibfnamefont{A.~A.} \bibnamefont{Belik}},
  \bibinfo{author}{\bibfnamefont{S.}~\bibnamefont{Iikubo}},
  \bibinfo{author}{\bibfnamefont{K.}~\bibnamefont{Kodama}},
  \bibinfo{author}{\bibfnamefont{N.}~\bibnamefont{Igawa}},
  \bibinfo{author}{\bibfnamefont{S.}~\bibnamefont{Shamoto}},
  \bibinfo{author}{\bibfnamefont{M.}~\bibnamefont{Maie}},
  \bibinfo{author}{\bibfnamefont{T.}~\bibnamefont{Nagai}},
  \bibinfo{author}{\bibfnamefont{Y.}~\bibnamefont{Matsui}},
  \bibinfo{author}{\bibfnamefont{S.~Y.} \bibnamefont{Stefanovich}},
  \bibinfo{author}{\bibfnamefont{B.~I.} \bibnamefont{Lazoryak}},
  \bibnamefont{et~al.}, \bibinfo{journal}{J. Am. Chem. Soc.}
  \textbf{\bibinfo{volume}{128}}, \bibinfo{pages}{706}
  (\bibinfo{year}{2006}{\natexlab{a}}).

\bibitem[{\citenamefont{Jia et~al.}(1995)\citenamefont{Jia, Callcott, Yurkas,
  Ellis, Himpsel, Samant, St{\"{o}}hr, Ederer, Carlisle, Hudson
  et~al.}}]{jia_95}
\bibinfo{author}{\bibfnamefont{J.~J.} \bibnamefont{Jia}},
  \bibinfo{author}{\bibfnamefont{T.~A.} \bibnamefont{Callcott}},
  \bibinfo{author}{\bibfnamefont{J.}~\bibnamefont{Yurkas}},
  \bibinfo{author}{\bibfnamefont{A.~W.} \bibnamefont{Ellis}},
  \bibinfo{author}{\bibfnamefont{F.~J.} \bibnamefont{Himpsel}},
  \bibinfo{author}{\bibfnamefont{M.~G.} \bibnamefont{Samant}},
  \bibinfo{author}{\bibfnamefont{J.}~\bibnamefont{St{\"{o}}hr}},
  \bibinfo{author}{\bibfnamefont{D.~L.} \bibnamefont{Ederer}},
  \bibinfo{author}{\bibfnamefont{J.~A.} \bibnamefont{Carlisle}},
  \bibinfo{author}{\bibfnamefont{E.~A.} \bibnamefont{Hudson}},
  \bibnamefont{et~al.}, \bibinfo{journal}{Rev. Sci. Instrum.}
  \textbf{\bibinfo{volume}{66}}, \bibinfo{pages}{1394} (\bibinfo{year}{1995}).

\bibitem[{\citenamefont{Regier et~al.}(2007)\citenamefont{Regier, Krochak,
  Sham, Hu, Thompson, and Blyth}}]{regier_07}
\bibinfo{author}{\bibfnamefont{T.}~\bibnamefont{Regier}},
  \bibinfo{author}{\bibfnamefont{J.}~\bibnamefont{Krochak}},
  \bibinfo{author}{\bibfnamefont{T.~K.} \bibnamefont{Sham}},
  \bibinfo{author}{\bibfnamefont{Y.~F.} \bibnamefont{Hu}},
  \bibinfo{author}{\bibfnamefont{J.}~\bibnamefont{Thompson}}, \bibnamefont{and}
  \bibinfo{author}{\bibfnamefont{R.~I.~R.} \bibnamefont{Blyth}},
  \bibinfo{journal}{Nucl. Instrum. Meth. A} \textbf{\bibinfo{volume}{582}},
  \bibinfo{pages}{93} (\bibinfo{year}{2007}).

\bibitem[{\citenamefont{Eitel et~al.}(2001)\citenamefont{Eitel, Randall,
  Shrout, Rehrig, Hackenberger, and Park}}]{bspt}
\bibinfo{author}{\bibfnamefont{R.~E.} \bibnamefont{Eitel}},
  \bibinfo{author}{\bibfnamefont{C.~A.} \bibnamefont{Randall}},
  \bibinfo{author}{\bibfnamefont{T.~R.} \bibnamefont{Shrout}},
  \bibinfo{author}{\bibfnamefont{P.~W.} \bibnamefont{Rehrig}},
  \bibinfo{author}{\bibfnamefont{W.}~\bibnamefont{Hackenberger}},
  \bibnamefont{and} \bibinfo{author}{\bibfnamefont{S.-E.} \bibnamefont{Park}},
  \bibinfo{journal}{Jpn. J. Appl. Phys.} \textbf{\bibinfo{volume}{40}},
  \bibinfo{pages}{5999} (\bibinfo{year}{2001}).

\bibitem[{\citenamefont{Sugawara et~al.}(1968)\citenamefont{Sugawara, Iiida,
  Syono, and Akimoto}}]{sugawara}
\bibinfo{author}{\bibfnamefont{F.}~\bibnamefont{Sugawara}},
  \bibinfo{author}{\bibfnamefont{S.}~\bibnamefont{Iiida}},
  \bibinfo{author}{\bibfnamefont{Y.}~\bibnamefont{Syono}}, \bibnamefont{and}
  \bibinfo{author}{\bibfnamefont{S.}~\bibnamefont{Akimoto}},
  \bibinfo{journal}{J. Phys. Soc. Jpn.} \textbf{\bibinfo{volume}{25}},
  \bibinfo{pages}{1553} (\bibinfo{year}{1968}).

\bibitem[{\citenamefont{Niitaka et~al.}(2004)\citenamefont{Niitaka, Azuma,
  Takano, Nishibori, Takata, and Sakata}}]{niitaka}
\bibinfo{author}{\bibfnamefont{S.}~\bibnamefont{Niitaka}},
  \bibinfo{author}{\bibfnamefont{M.}~\bibnamefont{Azuma}},
  \bibinfo{author}{\bibfnamefont{M.}~\bibnamefont{Takano}},
  \bibinfo{author}{\bibfnamefont{E.}~\bibnamefont{Nishibori}},
  \bibinfo{author}{\bibfnamefont{M.}~\bibnamefont{Takata}}, \bibnamefont{and}
  \bibinfo{author}{\bibfnamefont{M.}~\bibnamefont{Sakata}},
  \bibinfo{journal}{Solid State Ionics} \textbf{\bibinfo{volume}{172}},
  \bibinfo{pages}{557} (\bibinfo{year}{2004}).

\bibitem[{\citenamefont{Belik et~al.}(2007{\natexlab{b}})\citenamefont{Belik,
  Tsujii, Suzuki, and Takayama-Muromachi}}]{belik_07}
\bibinfo{author}{\bibfnamefont{A.~A.} \bibnamefont{Belik}},
  \bibinfo{author}{\bibfnamefont{N.}~\bibnamefont{Tsujii}},
  \bibinfo{author}{\bibfnamefont{H.}~\bibnamefont{Suzuki}}, \bibnamefont{and}
  \bibinfo{author}{\bibfnamefont{E.}~\bibnamefont{Takayama-Muromachi}},
  \bibinfo{journal}{Inorg. Chem.} \textbf{\bibinfo{volume}{46}},
  \bibinfo{pages}{8746} (\bibinfo{year}{2007}{\natexlab{b}}).

\bibitem[{\citenamefont{Montanari et~al.}(2007)\citenamefont{Montanari,
  Calestani, Righi, Gilioli, Bolzoni, Knight, and Radaelli}}]{montanari_07}
\bibinfo{author}{\bibfnamefont{E.}~\bibnamefont{Montanari}},
  \bibinfo{author}{\bibfnamefont{G.}~\bibnamefont{Calestani}},
  \bibinfo{author}{\bibfnamefont{L.}~\bibnamefont{Righi}},
  \bibinfo{author}{\bibfnamefont{E.}~\bibnamefont{Gilioli}},
  \bibinfo{author}{\bibfnamefont{F.}~\bibnamefont{Bolzoni}},
  \bibinfo{author}{\bibfnamefont{K.~S.} \bibnamefont{Knight}},
  \bibnamefont{and} \bibinfo{author}{\bibfnamefont{P.~G.}
  \bibnamefont{Radaelli}}, \bibinfo{journal}{Phys. Rev. B}
  \textbf{\bibinfo{volume}{75}}, \bibinfo{pages}{220101(R)}
  (\bibinfo{year}{2007}).

\bibitem[{\citenamefont{dos Santos et~al.}(2002{\natexlab{a}})\citenamefont{dos
  Santos, Cheetham, Atou, Syono, Yamaguchi, Ohoyama, Chiba, and
  Rao}}]{moreira_02}
\bibinfo{author}{\bibfnamefont{A.~M.} \bibnamefont{dos Santos}},
  \bibinfo{author}{\bibfnamefont{A.~K.} \bibnamefont{Cheetham}},
  \bibinfo{author}{\bibfnamefont{T.}~\bibnamefont{Atou}},
  \bibinfo{author}{\bibfnamefont{Y.}~\bibnamefont{Syono}},
  \bibinfo{author}{\bibfnamefont{Y.}~\bibnamefont{Yamaguchi}},
  \bibinfo{author}{\bibfnamefont{K.}~\bibnamefont{Ohoyama}},
  \bibinfo{author}{\bibfnamefont{H.}~\bibnamefont{Chiba}}, \bibnamefont{and}
  \bibinfo{author}{\bibfnamefont{C.~N.~R.} \bibnamefont{Rao}},
  \bibinfo{journal}{Phys. Rev. B} \textbf{\bibinfo{volume}{66}},
  \bibinfo{pages}{064425} (\bibinfo{year}{2002}{\natexlab{a}}).

\bibitem[{\citenamefont{Hill et~al.}(2002)\citenamefont{Hill, Battig, and
  Daul}}]{hill_02}
\bibinfo{author}{\bibfnamefont{N.~A.} \bibnamefont{Hill}},
  \bibinfo{author}{\bibfnamefont{P.}~\bibnamefont{Battig}}, \bibnamefont{and}
  \bibinfo{author}{\bibfnamefont{C.}~\bibnamefont{Daul}}, \bibinfo{journal}{J.
  Phys. Chem. B} \textbf{\bibinfo{volume}{106}}, \bibinfo{pages}{3383}
  (\bibinfo{year}{2002}).

\bibitem[{\citenamefont{Hill and Rabe}(1999)}]{HillRabe}
\bibinfo{author}{\bibfnamefont{N.~A.} \bibnamefont{Hill}} \bibnamefont{and}
  \bibinfo{author}{\bibfnamefont{K.~M.} \bibnamefont{Rabe}},
  \bibinfo{journal}{Phys. Rev. B} \textbf{\bibinfo{volume}{59}},
  \bibinfo{pages}{8759} (\bibinfo{year}{1999}).

\bibitem[{\citenamefont{Seshadri and Hill}(2001)}]{SeshadriHill}
\bibinfo{author}{\bibfnamefont{R.}~\bibnamefont{Seshadri}} \bibnamefont{and}
  \bibinfo{author}{\bibfnamefont{N.~A.} \bibnamefont{Hill}},
  \bibinfo{journal}{Chem. Mater.} \textbf{\bibinfo{volume}{13}},
  \bibinfo{pages}{2892} (\bibinfo{year}{2001}).

\bibitem[{\citenamefont{dos Santos et~al.}(2002{\natexlab{b}})\citenamefont{dos
  Santos, Parashar, Raju, Znao, Cheetham, and Rao}}]{SantosSSC}
\bibinfo{author}{\bibfnamefont{A.~M.} \bibnamefont{dos Santos}},
  \bibinfo{author}{\bibfnamefont{S.}~\bibnamefont{Parashar}},
  \bibinfo{author}{\bibfnamefont{A.~R.} \bibnamefont{Raju}},
  \bibinfo{author}{\bibfnamefont{Y.~S.} \bibnamefont{Znao}},
  \bibinfo{author}{\bibfnamefont{A.~K.} \bibnamefont{Cheetham}},
  \bibnamefont{and} \bibinfo{author}{\bibfnamefont{C.~N.~R.}
  \bibnamefont{Rao}}, \bibinfo{journal}{Solid State Commun.}
  \textbf{\bibinfo{volume}{122}}, \bibinfo{pages}{49}
  (\bibinfo{year}{2002}{\natexlab{b}}).

\bibitem[{\citenamefont{Shishidou et~al.}(2004)\citenamefont{Shishidou, Mikamo,
  Uratani, Ishii, and Oguchi}}]{Shishidou_04}
\bibinfo{author}{\bibfnamefont{T.}~\bibnamefont{Shishidou}},
  \bibinfo{author}{\bibfnamefont{N.}~\bibnamefont{Mikamo}},
  \bibinfo{author}{\bibfnamefont{Y.}~\bibnamefont{Uratani}},
  \bibinfo{author}{\bibfnamefont{F.}~\bibnamefont{Ishii}}, \bibnamefont{and}
  \bibinfo{author}{\bibfnamefont{T.}~\bibnamefont{Oguchi}},
  \bibinfo{journal}{J. Phys.: Condens. Matter} \textbf{\bibinfo{volume}{16}},
  \bibinfo{pages}{S5677} (\bibinfo{year}{2004}).

\bibitem[{\citenamefont{Atou et~al.}(1999)\citenamefont{Atou, Chiba, Ohoyama,
  Yamaguchi, and Syono}}]{atou_99}
\bibinfo{author}{\bibfnamefont{T.}~\bibnamefont{Atou}},
  \bibinfo{author}{\bibfnamefont{H.}~\bibnamefont{Chiba}},
  \bibinfo{author}{\bibfnamefont{K.}~\bibnamefont{Ohoyama}},
  \bibinfo{author}{\bibfnamefont{Y.}~\bibnamefont{Yamaguchi}},
  \bibnamefont{and} \bibinfo{author}{\bibfnamefont{Y.}~\bibnamefont{Syono}},
  \bibinfo{journal}{J. Solid State Chem.} \textbf{\bibinfo{volume}{145}},
  \bibinfo{pages}{639} (\bibinfo{year}{1999}).

\bibitem[{\citenamefont{dos Santos et~al.}(2002{\natexlab{c}})\citenamefont{dos
  Santos, Cheetham, Atou, Syono, Yamaguchi, Ohoyama, Chiba, and
  Rao}}]{santos_02}
\bibinfo{author}{\bibfnamefont{A.~M.} \bibnamefont{dos Santos}},
  \bibinfo{author}{\bibfnamefont{A.~K.} \bibnamefont{Cheetham}},
  \bibinfo{author}{\bibfnamefont{T.}~\bibnamefont{Atou}},
  \bibinfo{author}{\bibfnamefont{Y.}~\bibnamefont{Syono}},
  \bibinfo{author}{\bibfnamefont{Y.}~\bibnamefont{Yamaguchi}},
  \bibinfo{author}{\bibfnamefont{K.}~\bibnamefont{Ohoyama}},
  \bibinfo{author}{\bibfnamefont{H.}~\bibnamefont{Chiba}}, \bibnamefont{and}
  \bibinfo{author}{\bibfnamefont{C.~N.~R.} \bibnamefont{Rao}},
  \bibinfo{journal}{Phys. Rev. B} \textbf{\bibinfo{volume}{66}},
  \bibinfo{pages}{064425} (\bibinfo{year}{2002}{\natexlab{c}}).

\bibitem[{\citenamefont{Montanari et~al.}(2005)\citenamefont{Montanari,
  Calestani, Migliori, Dapiaggi, Bolzoni, Cabassi, and Gilioli}}]{montanari_05}
\bibinfo{author}{\bibfnamefont{E.}~\bibnamefont{Montanari}},
  \bibinfo{author}{\bibfnamefont{G.}~\bibnamefont{Calestani}},
  \bibinfo{author}{\bibfnamefont{A.}~\bibnamefont{Migliori}},
  \bibinfo{author}{\bibfnamefont{M.}~\bibnamefont{Dapiaggi}},
  \bibinfo{author}{\bibfnamefont{F.}~\bibnamefont{Bolzoni}},
  \bibinfo{author}{\bibfnamefont{R.}~\bibnamefont{Cabassi}}, \bibnamefont{and}
  \bibinfo{author}{\bibfnamefont{E.}~\bibnamefont{Gilioli}},
  \bibinfo{journal}{Chem. Mater.} \textbf{\bibinfo{volume}{17}},
  \bibinfo{pages}{6457} (\bibinfo{year}{2005}).

\bibitem[{\citenamefont{Shishidou}(2007)}]{Shishidou}
\bibinfo{author}{\bibfnamefont{T.}~\bibnamefont{Shishidou}}
  (\bibinfo{year}{2007}), \eprint{\textit{private communication}}.

\bibitem[{\citenamefont{Baettig et~al.}(2007)\citenamefont{Baettig, Seshadri,
  and Spaldin}}]{spaldin_07}
\bibinfo{author}{\bibfnamefont{P.}~\bibnamefont{Baettig}},
  \bibinfo{author}{\bibfnamefont{R.}~\bibnamefont{Seshadri}}, \bibnamefont{and}
  \bibinfo{author}{\bibfnamefont{N.~A.} \bibnamefont{Spaldin}},
  \bibinfo{journal}{J. Am. Chem. Soc.} \textbf{\bibinfo{volume}{129}},
  \bibinfo{pages}{9854} (\bibinfo{year}{2007}).

\bibitem[{\citenamefont{Solovyev and Pchelkina}(2008)}]{solovyev_08}
\bibinfo{author}{\bibfnamefont{I.~V.} \bibnamefont{Solovyev}} \bibnamefont{and}
  \bibinfo{author}{\bibfnamefont{Z.~V.} \bibnamefont{Pchelkina}},
  \bibinfo{journal}{New J. Phys.} \textbf{\bibinfo{volume}{10}},
  \bibinfo{pages}{073021} (\bibinfo{year}{2008}).

\bibitem[{\citenamefont{Solovyev and Pchelkina}(2009)}]{pzv_solovyev}
\bibinfo{author}{\bibfnamefont{I.~V.} \bibnamefont{Solovyev}} \bibnamefont{and}
  \bibinfo{author}{\bibfnamefont{Z.~V.} \bibnamefont{Pchelkina}},
  \bibinfo{journal}{Pis'ma v ZhETF.} \textbf{\bibinfo{volume}{89}},
  \bibinfo{pages}{701} (\bibinfo{year}{2009}).

\bibitem[{\citenamefont{Smolenskii and Chupis}(1982)}]{smolenskii_82}
\bibinfo{author}{\bibfnamefont{G.~A.} \bibnamefont{Smolenskii}}
  \bibnamefont{and} \bibinfo{author}{\bibfnamefont{I.}~\bibnamefont{Chupis}},
  \bibinfo{journal}{Sov. Phys. Usp.} \textbf{\bibinfo{volume}{25}},
  \bibinfo{pages}{475} (\bibinfo{year}{1982}).

\bibitem[{\citenamefont{Sosnowska et~al.}(2002)\citenamefont{Sosnowska,
  Sch{\"{a}}fer, Kockelmann, Andersen, and Troyanchuk}}]{sosnowska_02}
\bibinfo{author}{\bibfnamefont{I.}~\bibnamefont{Sosnowska}},
  \bibinfo{author}{\bibfnamefont{W.}~\bibnamefont{Sch{\"{a}}fer}},
  \bibinfo{author}{\bibfnamefont{W.}~\bibnamefont{Kockelmann}},
  \bibinfo{author}{\bibfnamefont{K.~H.} \bibnamefont{Andersen}},
  \bibnamefont{and} \bibinfo{author}{\bibfnamefont{I.~O.}
  \bibnamefont{Troyanchuk}}, \bibinfo{journal}{Appl. Phys. A}
  \textbf{\bibinfo{volume}{74}}, \bibinfo{pages}{S1040} (\bibinfo{year}{2002}).

\bibitem[{\citenamefont{Sosnowska et~al.}(1982)\citenamefont{Sosnowska,
  Peterlin-Neumaier, and Steichele}}]{sosnowska_82}
\bibinfo{author}{\bibfnamefont{I.}~\bibnamefont{Sosnowska}},
  \bibinfo{author}{\bibfnamefont{T.}~\bibnamefont{Peterlin-Neumaier}},
  \bibnamefont{and}
  \bibinfo{author}{\bibfnamefont{E.}~\bibnamefont{Steichele}},
  \bibinfo{journal}{J. Phys. C} \textbf{\bibinfo{volume}{15}},
  \bibinfo{pages}{4835} (\bibinfo{year}{1982}).

\bibitem[{\citenamefont{Teague et~al.}(1970)\citenamefont{Teague, Gerson, and
  James}}]{teague_70}
\bibinfo{author}{\bibfnamefont{J.~R.} \bibnamefont{Teague}},
  \bibinfo{author}{\bibfnamefont{R.}~\bibnamefont{Gerson}}, \bibnamefont{and}
  \bibinfo{author}{\bibfnamefont{W.~J.} \bibnamefont{James}},
  \bibinfo{journal}{Solid State Commun.} \textbf{\bibinfo{volume}{8}},
  \bibinfo{pages}{1073} (\bibinfo{year}{1970}).

\bibitem[{\citenamefont{Wang et~al.}(2003)\citenamefont{Wang, Neaton, Zheng,
  Nagarajan, Ogale, Liu, Viehland, Vaithyanathan, Schlom, Waghmare
  et~al.}}]{wang_03}
\bibinfo{author}{\bibfnamefont{J.}~\bibnamefont{Wang}},
  \bibinfo{author}{\bibfnamefont{J.~B.} \bibnamefont{Neaton}},
  \bibinfo{author}{\bibfnamefont{H.}~\bibnamefont{Zheng}},
  \bibinfo{author}{\bibfnamefont{V.}~\bibnamefont{Nagarajan}},
  \bibinfo{author}{\bibfnamefont{S.~B.} \bibnamefont{Ogale}},
  \bibinfo{author}{\bibfnamefont{B.}~\bibnamefont{Liu}},
  \bibinfo{author}{\bibfnamefont{D.}~\bibnamefont{Viehland}},
  \bibinfo{author}{\bibfnamefont{V.}~\bibnamefont{Vaithyanathan}},
  \bibinfo{author}{\bibfnamefont{D.~G.} \bibnamefont{Schlom}},
  \bibinfo{author}{\bibfnamefont{U.~V.} \bibnamefont{Waghmare}},
  \bibnamefont{et~al.}, \bibinfo{journal}{Science}
  \textbf{\bibinfo{volume}{299}}, \bibinfo{pages}{1719} (\bibinfo{year}{2003}).

\bibitem[{\citenamefont{Li et~al.}(2004)\citenamefont{Li, Wang, Wuttig, and
  Ramesh}}]{li_04}
\bibinfo{author}{\bibfnamefont{J.}~\bibnamefont{Li}},
  \bibinfo{author}{\bibfnamefont{J.}~\bibnamefont{Wang}},
  \bibinfo{author}{\bibfnamefont{M.}~\bibnamefont{Wuttig}}, \bibnamefont{and}
  \bibinfo{author}{\bibfnamefont{R.}~\bibnamefont{Ramesh}},
  \bibinfo{journal}{Appl. Phys. Lett.} \textbf{\bibinfo{volume}{84}},
  \bibinfo{pages}{5261} (\bibinfo{year}{2004}).

\bibitem[{\citenamefont{Yun et~al.}(2004)\citenamefont{Yun, Ricinschi,
  Kanashima, Noda, and Okuyama}}]{yun_04}
\bibinfo{author}{\bibfnamefont{K.~Y.} \bibnamefont{Yun}},
  \bibinfo{author}{\bibfnamefont{D.}~\bibnamefont{Ricinschi}},
  \bibinfo{author}{\bibfnamefont{T.}~\bibnamefont{Kanashima}},
  \bibinfo{author}{\bibfnamefont{M.}~\bibnamefont{Noda}}, \bibnamefont{and}
  \bibinfo{author}{\bibfnamefont{M.}~\bibnamefont{Okuyama}},
  \bibinfo{journal}{Jpn. J. Appl. Phys.} \textbf{\bibinfo{volume}{43}},
  \bibinfo{pages}{L647} (\bibinfo{year}{2004}).

\bibitem[{\citenamefont{King-Smith and Vanderbilt}(1993)}]{king_93}
\bibinfo{author}{\bibfnamefont{R.~D.} \bibnamefont{King-Smith}}
  \bibnamefont{and}
  \bibinfo{author}{\bibfnamefont{D.}~\bibnamefont{Vanderbilt}},
  \bibinfo{journal}{Phys. Rev. B} \textbf{\bibinfo{volume}{47}},
  \bibinfo{pages}{1651} (\bibinfo{year}{1993}).

\bibitem[{\citenamefont{Vanderbilt and King-Smith}(1993)}]{vanderbilt_93}
\bibinfo{author}{\bibfnamefont{D.}~\bibnamefont{Vanderbilt}} \bibnamefont{and}
  \bibinfo{author}{\bibfnamefont{R.~D.} \bibnamefont{King-Smith}},
  \bibinfo{journal}{Phys. Rev. B} \textbf{\bibinfo{volume}{48}},
  \bibinfo{pages}{4442} (\bibinfo{year}{1993}).

\bibitem[{\citenamefont{Restam}(1994)}]{resta_94}
\bibinfo{author}{\bibfnamefont{R.}~\bibnamefont{Restam}},
  \bibinfo{journal}{Rev. Mod. Phys.} \textbf{\bibinfo{volume}{66}},
  \bibinfo{pages}{899} (\bibinfo{year}{1994}).

\bibitem[{\citenamefont{Neaton et~al.}(2005)\citenamefont{Neaton, Ederer,
  Waghmare, Spaldin, and Rabe}}]{neaton_05}
\bibinfo{author}{\bibfnamefont{J.~B.} \bibnamefont{Neaton}},
  \bibinfo{author}{\bibfnamefont{C.}~\bibnamefont{Ederer}},
  \bibinfo{author}{\bibfnamefont{U.~V.} \bibnamefont{Waghmare}},
  \bibinfo{author}{\bibfnamefont{N.~A.} \bibnamefont{Spaldin}},
  \bibnamefont{and} \bibinfo{author}{\bibfnamefont{K.~M.} \bibnamefont{Rabe}},
  \bibinfo{journal}{Phys. Rev. B} \textbf{\bibinfo{volume}{71}},
  \bibinfo{pages}{014113} (\bibinfo{year}{2005}).

\bibitem[{\citenamefont{Belik et~al.}(2006{\natexlab{b}})\citenamefont{Belik,
  Iikubo, Kodama, Igawa, Shamoto, Niitaka, Azuma, Takano, Izumi, and
  Takayama-Muromachi}}]{belik_co_06}
\bibinfo{author}{\bibfnamefont{A.~A.} \bibnamefont{Belik}},
  \bibinfo{author}{\bibfnamefont{S.}~\bibnamefont{Iikubo}},
  \bibinfo{author}{\bibfnamefont{K.}~\bibnamefont{Kodama}},
  \bibinfo{author}{\bibfnamefont{N.}~\bibnamefont{Igawa}},
  \bibinfo{author}{\bibfnamefont{S.}~\bibnamefont{Shamoto}},
  \bibinfo{author}{\bibfnamefont{S.}~\bibnamefont{Niitaka}},
  \bibinfo{author}{\bibfnamefont{M.}~\bibnamefont{Azuma}},
  \bibinfo{author}{\bibfnamefont{M.}~\bibnamefont{Takano}},
  \bibinfo{author}{\bibfnamefont{F.}~\bibnamefont{Izumi}}, \bibnamefont{and}
  \bibinfo{author}{\bibfnamefont{E.}~\bibnamefont{Takayama-Muromachi}},
  \bibinfo{journal}{Chem. Mater.} \textbf{\bibinfo{volume}{18}},
  \bibinfo{pages}{798} (\bibinfo{year}{2006}{\natexlab{b}}).

\bibitem[{\citenamefont{Uratani et~al.}(2005)\citenamefont{Uratani, Shishidou,
  Ishii, and Oguchi}}]{uratani_05}
\bibinfo{author}{\bibfnamefont{Y.}~\bibnamefont{Uratani}},
  \bibinfo{author}{\bibfnamefont{T.}~\bibnamefont{Shishidou}},
  \bibinfo{author}{\bibfnamefont{F.}~\bibnamefont{Ishii}}, \bibnamefont{and}
  \bibinfo{author}{\bibfnamefont{T.}~\bibnamefont{Oguchi}},
  \bibinfo{journal}{Jpn. J. Appl. Phys.} \textbf{\bibinfo{volume}{44}},
  \bibinfo{pages}{7130} (\bibinfo{year}{2005}).

\bibitem[{\citenamefont{Ishiwata et~al.}(2002)\citenamefont{Ishiwata, Azuma,
  Takano, Nishibori, Takata, Sakata, and Kato}}]{ishiwata_02}
\bibinfo{author}{\bibfnamefont{S.}~\bibnamefont{Ishiwata}},
  \bibinfo{author}{\bibfnamefont{M.}~\bibnamefont{Azuma}},
  \bibinfo{author}{\bibfnamefont{M.}~\bibnamefont{Takano}},
  \bibinfo{author}{\bibfnamefont{E.}~\bibnamefont{Nishibori}},
  \bibinfo{author}{\bibfnamefont{M.}~\bibnamefont{Takata}},
  \bibinfo{author}{\bibfnamefont{M.}~\bibnamefont{Sakata}}, \bibnamefont{and}
  \bibinfo{author}{\bibfnamefont{K.}~\bibnamefont{Kato}}, \bibinfo{journal}{J.
  Mater. Chem.} \textbf{\bibinfo{volume}{12}}, \bibinfo{pages}{3733}
  (\bibinfo{year}{2002}).

\bibitem[{\citenamefont{Carlsson et~al.}(2008)\citenamefont{Carlsson, Azuma,
  Shimakawa, Takano, Hewat, and Attfield}}]{carlsson_08}
\bibinfo{author}{\bibfnamefont{S.~J.~E.} \bibnamefont{Carlsson}},
  \bibinfo{author}{\bibfnamefont{M.}~\bibnamefont{Azuma}},
  \bibinfo{author}{\bibfnamefont{Y.}~\bibnamefont{Shimakawa}},
  \bibinfo{author}{\bibfnamefont{M.}~\bibnamefont{Takano}},
  \bibinfo{author}{\bibfnamefont{A.}~\bibnamefont{Hewat}}, \bibnamefont{and}
  \bibinfo{author}{\bibfnamefont{J.~P.} \bibnamefont{Attfield}},
  \bibinfo{journal}{J. Solid State Chem.} \textbf{\bibinfo{volume}{181}},
  \bibinfo{pages}{611} (\bibinfo{year}{2008}).

\bibitem[{\citenamefont{Azuma et~al.}(2007)\citenamefont{Azuma, Carlsson,
  Rodgers, Tucker, Tsujimoto, Ishiwata, Isoda, Shimakawa, Takano, and
  Attfield}}]{azuma_07}
\bibinfo{author}{\bibfnamefont{M.}~\bibnamefont{Azuma}},
  \bibinfo{author}{\bibfnamefont{S.}~\bibnamefont{Carlsson}},
  \bibinfo{author}{\bibfnamefont{J.}~\bibnamefont{Rodgers}},
  \bibinfo{author}{\bibfnamefont{M.~G.} \bibnamefont{Tucker}},
  \bibinfo{author}{\bibfnamefont{M.}~\bibnamefont{Tsujimoto}},
  \bibinfo{author}{\bibfnamefont{S.}~\bibnamefont{Ishiwata}},
  \bibinfo{author}{\bibfnamefont{S.}~\bibnamefont{Isoda}},
  \bibinfo{author}{\bibfnamefont{Y.}~\bibnamefont{Shimakawa}},
  \bibinfo{author}{\bibfnamefont{M.}~\bibnamefont{Takano}}, \bibnamefont{and}
  \bibinfo{author}{\bibfnamefont{J.~P.} \bibnamefont{Attfield}},
  \bibinfo{journal}{J. Am. Chem. Soc.} \textbf{\bibinfo{volume}{129}},
  \bibinfo{pages}{14433} (\bibinfo{year}{2007}).

\bibitem[{\citenamefont{Kohn and Sham}(1965)}]{lda}
\bibinfo{author}{\bibfnamefont{W.}~\bibnamefont{Kohn}} \bibnamefont{and}
  \bibinfo{author}{\bibfnamefont{L.~J.} \bibnamefont{Sham}},
  \bibinfo{journal}{Phys. Rev.} \textbf{\bibinfo{volume}{140}},
  \bibinfo{pages}{A1133} (\bibinfo{year}{1965}).

\bibitem[{\citenamefont{Hedin and Lundqvist}(1971)}]{lda_1}
\bibinfo{author}{\bibfnamefont{L.}~\bibnamefont{Hedin}} \bibnamefont{and}
  \bibinfo{author}{\bibfnamefont{B.~I.} \bibnamefont{Lundqvist}},
  \bibinfo{journal}{J. Phys. C} \textbf{\bibinfo{volume}{4}},
  \bibinfo{pages}{2064} (\bibinfo{year}{1971}).

\bibitem[{\citenamefont{Andersen}(1975)}]{lmto}
\bibinfo{author}{\bibfnamefont{O.~K.} \bibnamefont{Andersen}},
  \bibinfo{journal}{Phys. Rev. B} \textbf{\bibinfo{volume}{12}},
  \bibinfo{pages}{3060} (\bibinfo{year}{1975}).

\bibitem[{\citenamefont{Anisimov et~al.}(1997)\citenamefont{Anisimov,
  Aryasetiawan, and Lichtenstein}}]{lsda_u}
\bibinfo{author}{\bibfnamefont{V.~I.} \bibnamefont{Anisimov}},
  \bibinfo{author}{\bibfnamefont{F.}~\bibnamefont{Aryasetiawan}},
  \bibnamefont{and} \bibinfo{author}{\bibfnamefont{A.~I.}
  \bibnamefont{Lichtenstein}}, \bibinfo{journal}{J. Phys.: Condens. Matter}
  \textbf{\bibinfo{volume}{9}}, \bibinfo{pages}{767} (\bibinfo{year}{1997}).

\bibitem[{\citenamefont{Gunnarsson et~al.}(1989)\citenamefont{Gunnarsson,
  Andersen, Jepsen, and Zaanen}}]{constr}
\bibinfo{author}{\bibfnamefont{O.}~\bibnamefont{Gunnarsson}},
  \bibinfo{author}{\bibfnamefont{O.~K.} \bibnamefont{Andersen}},
  \bibinfo{author}{\bibfnamefont{O.}~\bibnamefont{Jepsen}}, \bibnamefont{and}
  \bibinfo{author}{\bibfnamefont{J.}~\bibnamefont{Zaanen}},
  \bibinfo{journal}{Phys. Rev. B} \textbf{\bibinfo{volume}{39}},
  \bibinfo{pages}{1708} (\bibinfo{year}{1989}).

\bibitem[{\citenamefont{Kimura et~al.}(2003)\citenamefont{Kimura, Kawamoto,
  Yamada, Azuma, Takano, and Tokura}}]{tokura_03}
\bibinfo{author}{\bibfnamefont{T.}~\bibnamefont{Kimura}},
  \bibinfo{author}{\bibfnamefont{S.}~\bibnamefont{Kawamoto}},
  \bibinfo{author}{\bibfnamefont{I.}~\bibnamefont{Yamada}},
  \bibinfo{author}{\bibfnamefont{M.}~\bibnamefont{Azuma}},
  \bibinfo{author}{\bibfnamefont{M.}~\bibnamefont{Takano}}, \bibnamefont{and}
  \bibinfo{author}{\bibfnamefont{Y.}~\bibnamefont{Tokura}},
  \bibinfo{journal}{Phys. Rev. B} \textbf{\bibinfo{volume}{67}},
  \bibinfo{pages}{180401} (\bibinfo{year}{2003}).

\bibitem[{\citenamefont{Gao et~al.}(2006)\citenamefont{Gao, Yuan, Wang, Chen,
  Chen, Liu, and Ren}}]{gao_06}
\bibinfo{author}{\bibfnamefont{F.}~\bibnamefont{Gao}},
  \bibinfo{author}{\bibfnamefont{Y.}~\bibnamefont{Yuan}},
  \bibinfo{author}{\bibfnamefont{K.~F.} \bibnamefont{Wang}},
  \bibinfo{author}{\bibfnamefont{X.~Y.} \bibnamefont{Chen}},
  \bibinfo{author}{\bibfnamefont{F.}~\bibnamefont{Chen}},
  \bibinfo{author}{\bibfnamefont{J.-M.} \bibnamefont{Liu}}, \bibnamefont{and}
  \bibinfo{author}{\bibfnamefont{Z.~F.} \bibnamefont{Ren}},
  \bibinfo{journal}{Appl. Phys. Lett.} \textbf{\bibinfo{volume}{89}},
  \bibinfo{pages}{102506} (\bibinfo{year}{2006}).

\bibitem[{\citenamefont{Kurmaev et~al.}(2008)\citenamefont{Kurmaev, Wilks,
  Moewes, Finkelstein, Shamin, and Kune{\v{s}}}}]{kurmaev_08}
\bibinfo{author}{\bibfnamefont{E.~Z.} \bibnamefont{Kurmaev}},
  \bibinfo{author}{\bibfnamefont{R.~G.} \bibnamefont{Wilks}},
  \bibinfo{author}{\bibfnamefont{A.}~\bibnamefont{Moewes}},
  \bibinfo{author}{\bibfnamefont{L.~D.} \bibnamefont{Finkelstein}},
  \bibinfo{author}{\bibfnamefont{S.~N.} \bibnamefont{Shamin}},
  \bibnamefont{and}
  \bibinfo{author}{\bibfnamefont{J.}~\bibnamefont{Kune{\v{s}}}},
  \textbf{\bibinfo{volume}{77}}, \bibinfo{pages}{Phys. Rev. B}
  (\bibinfo{year}{2008}).

\bibitem[{\citenamefont{Kharel et~al.}(2009)\citenamefont{Kharel, Talebi,
  Ramachandran, Dixit, Naik, Sahana, Naik, Rao, and Lawes}}]{kharel_09}
\bibinfo{author}{\bibfnamefont{P.}~\bibnamefont{Kharel}},
  \bibinfo{author}{\bibfnamefont{S.}~\bibnamefont{Talebi}},
  \bibinfo{author}{\bibfnamefont{B.}~\bibnamefont{Ramachandran}},
  \bibinfo{author}{\bibfnamefont{A.}~\bibnamefont{Dixit}},
  \bibinfo{author}{\bibfnamefont{V.~M.} \bibnamefont{Naik}},
  \bibinfo{author}{\bibfnamefont{M.~B.} \bibnamefont{Sahana}},
  \bibinfo{author}{\bibfnamefont{R.}~\bibnamefont{Naik}},
  \bibinfo{author}{\bibfnamefont{M.~S.~R.} \bibnamefont{Rao}},
  \bibnamefont{and} \bibinfo{author}{\bibfnamefont{G.}~\bibnamefont{Lawes}},
  \bibinfo{journal}{J. Phys.: Condensed Matter} \textbf{\bibinfo{volume}{21}},
  \bibinfo{pages}{036001} (\bibinfo{year}{2009}).

\bibitem[{\citenamefont{McLeod et~al.}(2009)\citenamefont{McLeod, Wilks,
  Skorikov, Finkelstein, Abu-Samak, Kurmaev, and Moewes}}]{skorikov}
\bibinfo{author}{\bibfnamefont{J.~A.} \bibnamefont{McLeod}},
  \bibinfo{author}{\bibfnamefont{R.~G.} \bibnamefont{Wilks}},
  \bibinfo{author}{\bibfnamefont{N.~A.} \bibnamefont{Skorikov}},
  \bibinfo{author}{\bibfnamefont{L.~D.} \bibnamefont{Finkelstein}},
  \bibinfo{author}{\bibfnamefont{M.}~\bibnamefont{Abu-Samak}},
  \bibinfo{author}{\bibfnamefont{E.~Z.} \bibnamefont{Kurmaev}},
  \bibnamefont{and} \bibinfo{author}{\bibfnamefont{A.}~\bibnamefont{Moewes}}
  (\bibinfo{year}{2009}), \eprint{cond-mat/0908.1581v1}.

\bibitem[{\citenamefont{Kuiper et~al.}(1989)\citenamefont{Kuiper, Kruizinga,
  Ghijsen, Sawatzky, and Verweij}}]{kuiper_89}
\bibinfo{author}{\bibfnamefont{P.}~\bibnamefont{Kuiper}},
  \bibinfo{author}{\bibfnamefont{G.}~\bibnamefont{Kruizinga}},
  \bibinfo{author}{\bibfnamefont{J.}~\bibnamefont{Ghijsen}},
  \bibinfo{author}{\bibfnamefont{G.~A.} \bibnamefont{Sawatzky}},
  \bibnamefont{and} \bibinfo{author}{\bibfnamefont{H.}~\bibnamefont{Verweij}},
  \bibinfo{journal}{Phys. Rev. Lett.} \textbf{\bibinfo{volume}{62}},
  \bibinfo{pages}{221} (\bibinfo{year}{1989}).

\bibitem[{\citenamefont{Galakhov et~al.}(2002)\citenamefont{Galakhov, Karelina,
  Kellerman, Gorshkov, Ovechkina, and Neumann}}]{galakhov_02}
\bibinfo{author}{\bibfnamefont{V.~R.} \bibnamefont{Galakhov}},
  \bibinfo{author}{\bibfnamefont{V.~V.} \bibnamefont{Karelina}},
  \bibinfo{author}{\bibfnamefont{D.~G.} \bibnamefont{Kellerman}},
  \bibinfo{author}{\bibfnamefont{V.~S.} \bibnamefont{Gorshkov}},
  \bibinfo{author}{\bibfnamefont{N.~A.} \bibnamefont{Ovechkina}},
  \bibnamefont{and} \bibinfo{author}{\bibfnamefont{M.}~\bibnamefont{Neumann}},
  \bibinfo{journal}{Phys. Solid State} \textbf{\bibinfo{volume}{44}},
  \bibinfo{pages}{266} (\bibinfo{year}{2002}).

\bibitem[{\citenamefont{Abbate et~al.}(2002)\citenamefont{Abbate, Zampieri,
  Prado, Caneiro, Gonzalez-Calbet, and Vallet-Regi}}]{abbate_02}
\bibinfo{author}{\bibfnamefont{M.}~\bibnamefont{Abbate}},
  \bibinfo{author}{\bibfnamefont{G.}~\bibnamefont{Zampieri}},
  \bibinfo{author}{\bibfnamefont{F.}~\bibnamefont{Prado}},
  \bibinfo{author}{\bibfnamefont{A.}~\bibnamefont{Caneiro}},
  \bibinfo{author}{\bibfnamefont{J.~M.} \bibnamefont{Gonzalez-Calbet}},
  \bibnamefont{and}
  \bibinfo{author}{\bibfnamefont{M.}~\bibnamefont{Vallet-Regi}},
  \bibinfo{journal}{Phys. Rev. B} \textbf{\bibinfo{volume}{65}},
  \bibinfo{pages}{155101} (\bibinfo{year}{2002}).

\bibitem[{\citenamefont{van~den Brink and Khomskii}(2008)}]{brink_08}
\bibinfo{author}{\bibfnamefont{J.}~\bibnamefont{van~den Brink}}
  \bibnamefont{and} \bibinfo{author}{\bibfnamefont{D.~I.}
  \bibnamefont{Khomskii}}, \bibinfo{journal}{J. Phys.: Condensed Matter}
  \textbf{\bibinfo{volume}{20}}, \bibinfo{pages}{434217}
  (\bibinfo{year}{2008}).

\end{thebibliography}
\end{document}